\documentclass[12pt,manuscript]{aastex}

\slugcomment{Not to appear in Nonlearned J., 45.}

\shorttitle{The edge of the Galactic disc}
\shortauthors{Carraro et al.}

\begin{document}


\title{The edge of the young Galactic disc}


\author{Giovanni Carraro\altaffilmark{1}}
\affil{ESO, Alonso de Cordova 3107, 19100 Santiago de Chile, Chile}

\author{Ruben A. V\'azquez}
\affil{Facultad de Ciencias Astron\'omicas y Geof\'{\i}sicas (UNLP),
                 Instituto de Astrof\'{\i}sica de La Plata (CONICET, UNLP), \\
                 \hspace{0.2cm}Paseo del Bosque s/n, La Plata, Argentina}

\author{Edgardo Costa}
\affil{Departamento de Astronom\'ia, Universidad de Chile,
                 Casilla 36-D, Santiago, Chile}

\author{Gabriel Perren\altaffilmark{2}}
\affil{Instituto de F\'{\i}sica de Rosario, IFIR (CONICET - UNR), Parque Urquiza - 2000 Rosario - Argentina}

\and

\author{Andr\'e Moitinho}
\affil{SIM/IDL, Faculdade de Ci\'encias da Universidade de Lisboa, Ed. C8, Campo Grande, 1749-016 Lisboa, Portugal}


\altaffiltext{1}{On leave from Dipartimento di Astronomia, Universit\'a di 
Padova, Vicolo Osservatorio 5, I-35122, Padova, Italy}
\altaffiltext{2}{G. Perren is fellow of the CONICET, Argentina}
        

\begin{abstract}

In this work we report and discuss the detection of two distant diffuse stellar groups in the 
third Galactic quadrant. They are composed of young stars, with spectral types ranging 
from late O to late B, and lie at galactocentric distances between 15 and 20 kpc.
These groups are located in the area of two cataloged open clusters (VdB-Hagen~04 and
Ruprecht~30), projected towards the Vela-Puppis constellations, and within the core
of the Canis Major over-density. Their reddening and distance has been estimated
analyzing their color-color and color-magnitude diagrams, derived from deep $UBV$ 
photometry. The existence of young star aggregates at such extreme distances from the
Galactic center challenges the commonly accepted scenario in which the Galactic disc has
a sharp cut-off at about 14 kpc from the Galactic center, and indicates that it extends
to much greater distances (as also supported by recent detection of CO molecular
complexes well beyond this distance). While the groups we find in the area of Ruprecht~30 
are compatible with the Orion and
Norma-Cygnus spiral arms, respectively, the distant group we identify in the region of
VdB-Hagen~4 lies in the external regions of the Norma-Cygnus arm, at a galactocentric
distance ($\sim$20 kpc) where no young stars had been detected so far in the optical.

\end{abstract}





\section{Introduction}
Although the existence of a conspicuous extinction window in the direction of the third
Galactic quadrant (3GQ) has been known for decades (see e.g. Fitzgerald 1968, Moffat et
al. 1979, Janes 1991, Moitinho 2001), only recently it has been fully exploited to probe the structure 
of the outer Galactic disc in the optical domain. Major results from this observational
effort have been the detection of (1) previously unknown spiral features (Carraro et al.
2005, Moitinho et al. 2006, V\'azquez et al. 2008) and (2) of stellar over-densities: the Canis Major - CMa
over-density (Martin et al. 2004) and the Monoceros Ring - MRi (Newberg et al. 2002),
both believed to be signatures of past accretion events.\\

\noindent
Our research group has contributed substantially to the subject, providing a new
picture of the outer disc spiral structure. Briefly, the most important findings are:
(1) the outer (Norma-Cygnus) arm has been found to be a grand design spiral feature
defined by young stars, (2) the region closer to the Sun, at galactocentric distances
smaller than 9 kpc, is dominated by a conspicuous inter-arm structure, the Local
(Orion) spiral arm, at l $\sim$ 245$^{o}$, (3) the Perseus arm seems to be defined
in the 3GQ by gas and dust, and does not appear to be traced by an evident (optical)
young stellar population.\\

\noindent
Here we report the detection of two diffuse stellar groups in the 3GQ,
containing young stars with spectral types ranging from late O to A0.5, that lie
at galactocentric distances between 15 and 20 kpc, beyond the widely accepted radius
of the Galactic disk (14 kpc). The groups are located in the direction of two cataloged
open clusters (VdB-Hagen~04 and Ruprecht~30, Dias et al. 2002), projected towards the Vela-Puppis
constellations, well within the core of the Canis Major over-density.\\

\noindent
This is the fist time that such  young population is detected in the very outer disk
in optical. HI in our Galaxy extends out to galactocentric distances of 25 kpc, and
CO clouds have been found out to 20 kpc (Brand \& Wouterloot 2007). In the Infra-red (IR), compact and well-confined
regions of star formation has been recently detected by various group
(Snell et al. 2002; Yun et al. 2007, 2009; Brand \& Wouterloot 2007, both in the second and
in the third Galactic quadrant.

\noindent
The existence of young star aggregates at such extreme distances from the Galactic center
defies the commonly accepted scenario in which the Galactic disc has a sharp cut-off at
about 14 kpc from the Galactic center (Robin \& Cr\'ez\'e 1986ab, Robin et al. 1992), and
indicates that it extends to much greater distances. The meaning of this cut-off has been
questioned by Momany et al. (2006), as far as old/intermediate age populations are
concerned. We recall also that the absence of structures beyond this distance
in model Color Magnitude Diagrams  has been erroneously used as an evidence
to support the existence of the Canis Major dwarf Galaxy (Martin et al. 2004). 
Quite recently Sale et al. (2009) make use of the IPHAS surveys to address in a statistical
way the same point. By using over 40,000 A type stars they argue that they do not see any
abrupt truncation of the stellar density profile.

\noindent
Along the same vein, but with the greater details provided by the use of multicolor photometry,
we study in this work the sparse young population
in the (far-) outer disk along two lines of sight, 
and show that the stellar density profile of young stars (earlier than
A0) does not drop suddenly at 12-14 kpc from the Galactic center, but smoothly declines out
to 20 kpc. Together with the young, sparse population, we confirm the existence and properties of a compact,
extremely distant star cluster, VdB-Hagen 04, located at more than 20 kpc from the Galactic center.

\section{Observations and Data Reduction}

\subsection{Observations}       

The regions of interest (see Fig.~1) were observed with the Y4KCAM camera attached to the Cerro
Tololo Inter-American Observatory (CTIO) 1.0m telescope, operated by the SMARTS consortium
\footnote{\tt http://http://www.astro.yale.edu/smarts}. This camera is equipped with an STA
4064$\times$4064 CCD with 15-$\mu$ pixels, yielding a scale of 0.289$^{\prime\prime}$/pixel
and a field-of-view (FOV) of $20^{\prime} \times 20^{\prime}$ at the
Cassegrain focus of the CTIO 1.0m telescope. The CCD was operated without binning, at a nominal
gain of 1.44 e$^-$/ADU, implying a readout noise of 7~e$^-$ per quadrant (this detector is read
by means of four different amplifiers). QE and other detector characteristics can be found at:
http://www.astronomy.ohio-state.edu/Y4KCam/detector.html.\\

\noindent
In Table~1 we present the log of $UBV$ observations. All observations were carried out in
photometric, good seeing, conditions. Typical values for the seeing were 1.1, 1.4, 1.5 and
1.0 for January 29, January 31, February 1, and February 3, 2008, respectively.
Our $UBV$ instrumental photometric system was defined
by the use of a standard broad-band Kitt Peak $UBV$ set. Transmission curves
for these filters can be found at: http://www.astronomy.ohio-state.edu/Y4KCam/filters.html.
To determine the transformation from our instrumental system to the standard Johnson-Kron-Cousins
system, and to correct for extinction, we observed 46 stars in area SA~98 (Landolt 1992)
multiple times, and with different
airmasses ranging from $\sim$1.1 to $\sim$2.6.  Field SA~98 is very advantageous, as it
includes a large number of well observed standard stars, and it is completely covered by
the CCD's FOV.  Furthermore, the standard's color coverage is very good, being:
$-0.2 \leq (B-V)\leq 2.2$ and $-0.1 \leq (V-I) \leq 6.0$.\\

\subsection{Reductions}

Basic calibration of the CCD frames was done using the Yale/SMARTS y4k reduction script
based on the IRAF\footnote{IRAF is distributed
by the National Optical Astronomy Observatory, which is operated by the Association
of Universities for Research in Astronomy, Inc., under cooperative agreement with
the National Science Foundation.} package CCDRED. For this purpose, zero exposure
frames and twilight sky flats were taken every night.  Photometry was then performed
using the IRAF DAOPHOT and PHOTCAL packages. Instrumental magnitudes were extracted
following the Point Spread Function (PSF) method (Stetson 1987). A quadratic, spatially
variable, Master PSF (PENNY function) was adopted. Aperture corrections were determined
making aperture photometry of a suitable number (typically 20 to 40) of bright, isolated,
stars in the field. These corrections were found to vary from 0.160 to 0.290 mag, depending
on the seeing and filter. The PSF photometry was finally aperture corrected, filter by filter.

\subsection{The photometry}

Both fields were observed in two different nights, all four photometric.
We decided to shift observations to a single night for each field,
namely February 3 for VdB-Hagen~04, and January 31 for Ruprecht~30, since
these two nights had better seeing.

After removing problematic stars, and stars having only a few observations in Landolt's
(1992) catalog, our photometric solution for a grand total of 327 measurements per filter
- obtained by combining standard star observations from all nights - turned out to be:\\

\noindent
$ U = u + (3.097\pm0.010) + (0.44\pm0.01) \times X - (0.040\pm0.006) \times (U-B)$ \\
$ B = b + (2.103\pm0.012) + (0.27\pm0.01) \times X - (0.120\pm0.007) \times (B-V)$ \\
$ V = v + (1.760\pm0.007) + (0.14\pm0.01) \times X + (0.022\pm0.007) \times (B-V)$ \\

\noindent
for January 31, and

\noindent
$ U = u + (3.090\pm0.010) + (0.45\pm0.01) \times X - (0.040\pm0.006) \times (U-B)$ \\
$ B = b + (2.107\pm0.012) + (0.25\pm0.01) \times X - (0.111\pm0.007) \times (B-V)$ \\
$ V = v + (1.757\pm0.007) + (0.15\pm0.01) \times X + (0.018\pm0.007) \times (B-V)$ \\

\noindent
for February 3, respectively.\\
The final {\it r.m.s} of the fitting in both cases was 0.050, 0.030 and 0.020 in $U$, $B$,
and $V$, respectively.

\noindent
Global photometric errors were estimated using the scheme developed by Patat \& Carraro
(2001, Appendix A1), which takes into account the errors resulting from the PSF fitting
procedure (e.i. from ALLSTAR), and the calibration errors (corresponding to the zero point,
color terms and extinction errors). In Fig.~2 we present global photometric error trends 
plotted as a function of $V$ magnitude. Quick inspection shows that stars brighter than
$V \approx 20$ mag have errors lower than 0.05~mag in magnitude and lower than 0.10~mag in
all colors. \\

Our final optical photometric catalogs consist of 6039 entries for Ruprecht~30 and
3957 for VdB-Hagen~04 having $UBV$ measures down
to $V \sim $ 20.

Our optical catalogue was cross-correlated with 2MASS (Skrutskie et al. 2006), which
resulted in a final catalog including $UBV$ and $JHK_{s}$ magnitudes. As a by product,
pixel (detector) coordinates were converted to RA and DEC for J2000.0 equinox, thus
providing 2MASS-based astrometry.\\

\noindent
Finally, completeness corrections were determined by running artificial star experiments
on the data. In brief, we created several artificial images by adding artificial stars
to the original frames. These stars were added at random positions, and had the same
color and luminosity distribution of the true sample. To avoid generating over-crowding,
in each experiment we added up to 20\% of the original number of stars. Depending on
the frame, between 1000-5000 stars were added. In this way we have estimated that the
completeness level of our photometry is better than 50\% down to $V$ = 20.5.\\

\noindent
\noindent
The two fields in Fig.~1 are centered on cataloged Galactic clusters,
VdB-Hagen~04 (van den Bergh and Hagen 1976, Carraro and Costa 2007), and Ruprecht~30 ((Ruprecht 1966).
VdB-Hagen~04
 is a compact young cluster whose distance was earlier estimated to be larger than 19.0 kpc
from the Sun (Carraro \& Costa 2007), basing on just V and I photometry. This small color
coverage did not allow to estimate in a solid and precise way reddening and distance, since we could only rely
on the comparison with isochrones. By performing star counts with the new data-set described in this paper
we can confirm that VdB-Hagen is indeed an obvious  compact cluster (left panel of Fig.~3), with a radius of about
1.0 arcmin (right panel of Fig.~3). Here the radius is considered as the distance from the cluster center
at which star counts flatten down to the field level.
As for Ruprecht~30, we carried out a similar analysis. Star counts performed in this field do not
reveal any obvious overdensity (see Fig.~4), demonstrating the there is not cluster at the
location of Ruprecht 30.\\

\section{Young groups of OB stars in the extreme periphery of the Galactic disc: detection method}

The technique we used to extract information on the stellar populations present in the fields
from $UBV$ photometry is old and well established. It combines color-magnitude (CMD) and
two-color diagrams (TCD), and its success depends mainly on the availability 
of $U$-band photometry. A classical description of this method is given by Straizys (1995),
and recent applications of the procedure are illustrated in V\'azquez et al. (2005), Carraro et al. (2007), 
Carraro \& Costa (2009) and
V\'azquez et al. (2010). 
Briefly, the starting point is to construct $V$ {\it vs.} $(B-V)$ CMDs
(which as a standalone are difficult to interpret because we are dealing simultaneously with
age, distance, reddening and metallicity effects), and then to construct $(U-B)$ {\it vs.}
$(B-V)$ TCDs in which young blue stars of spectral type earlier than A0.5 immediately stand out
in the upper left part of the diagram, and then analyze the TCD for the stars contained in each strip.
As already demonstrated in previous papers (e.g. the case of the old Galactic cluster Auner~1 in Carraro et al. 2007), 
a powerful technique
is to cut the CMD in strips 1 mag wide, and then analyse the TCD for the stars contained
in each strip.\\
The CMDs are shown in Fig.~5 for Field~1 (VdB-Hagen~04, left panel), and Field~2 (Ruprecht~30, right panel),
where we indicate the location of Reg Giants and Blue Plume stars explicitly, to guide the reader.
Here only stars having simultaneously U, B and V measures with photometric errors smaller than 0.10 mag 
are shown.
The same stars are then plotted in the various panels of Fig.~4 for VdB-Hagen~04 and Fig.~5 for
Ruprecht~30.
In each of the ten panels in Figs.~6 and ~7, dashed lines indicate the run of interstellar reddening
for a few spectral type stars, to guide the eye.  They are drawn adopting
a normal  reddening law ($R_V$=3.1), which has been proved valid for the 3GQ ($R_V$=3.1, Moitinho 2001).
The solid line, on the other hand, is an empirical reddening-free, solar metallicity, 
Zero Age Main Sequence (ZAMS) from Schmidt-Kaler
(1982). We would like to stress however that in the region of the TCD we are interested to,
metallicity effects are
negligible, as amply discussed in Carraro et al. (2008).\\
The advantage of this CMD segmentation
is that more distant, hence more reddened, early type stars immediately stand out, and
one can easily separate groups of OB stars located at different distances.\\

\noindent
To determine reddening, spectral type and photometric distance we then proceed
as follows.
First we derive intrinsic colors using the two relationships:

\begin{equation}
E(U-B) = 0.72 \times E(B-V) + 0.05 \times E(B-V)^{2} ,
\end{equation}

\noindent
and

\begin{equation}
(U-B)_0 = 3.69 \times (B-V)_0 + 0.03.
\end{equation}

\noindent
The intrinsic color (B-V)$_0$ is the positive root of the second order equation one
derive combines the above expressions.
Intrinsic colors  ((U-B)$_0$ and (B-V)$_0$) are then directly 
correlated to spectral type, as compiled for instance in
Schmidt-Kaler (1982). The solution
of the equations above therefore allows us to encounter stars having spectral types
earlier than A0.5. For these stars we then know the absolute magnitude M$_{V}$ (again from the
Schmidt-Kaler 1982 compilation)
and, from 
the apparent extinction-corrected magnitude V$_{0}$, we finally infer the photometric distance.\\
Error in distances are computed as follows:

\begin{description}
\item $\Delta$ (Dist)  = ln(10) $\times$ Dist $\times$ $\Delta$ [log(Dist)];
\item $\Delta$ [log(Dist)] = $\frac{1}{5}$ $\times$ $\Delta$ V + $\Delta$ (M$_V$) + $\Delta$ (A$_V$)];
\item $\Delta$ (M$_V$) = 0;
\item $\Delta$ (A$_V$) = 3.1 $\times$ $\Delta$ (B-V);
\item $\Delta$ (V) and $\Delta$(B-V)  directly comes from photometry; finally
\item $\Delta$ (Dist) = ln(10) $\times$ Dist $\times$ 1/5 $\times$ [ $\Delta$ V + 3.1$\times$ $\Delta$ (B-V) ]
\end{description}

\noindent
Th results are then summarized in Tables 3 and 4, where we report for any detected early type  star, its ID,
magnitude, colors, reddening-corrected colors and magnitude, estimated spectral type and, finally,
the heliocentric distance with the associated uncertainty.\\
\noindent
In the following Sections~4 and 5 we are going to present a qualitative analysis
of the result for Field~1 and Field~2, respectively. A quantitative analysis of the early
star distributions in the fields is deferred to Section~6.

\section{The field toward VdB-Hagen~04}
\noindent
{\bf Field 1}:

This field is centered on the Galactic star cluster VdB-Hagen 04.\\

\noindent
The overall star distribution seen in the CMD of Field 1 (left panel of Fig.~5) is typical of stellar fields in the
3GQ (Moitinho 2001, Moitinho et al. 2006, Carraro et al. 2007, 2008). 
Apart from the  obvious main sequence (MS) produced by nearby
stars, the basic features of its CMD are: (1) a prominent
bright blue sequence, commonly referred to as {\it blue plume}, which is the target of this
study; (2) a thick blue MS downward of $V$ $\sim$ 18; and (3) a population of red giants, showing
a significant spread in both color and magnitude.\\

\noindent
Moving now to Fig.~6, where different TCDs are shown as a function of the magnitude V, the following
remarks can be done:

\begin{description}
\item $\bullet$ no stars earlier than A0 can be found for V brighter than 14.0 mag;
\item $\bullet$ only at V = 15.0 the first stars of early spectral type start to appear,
and become more and more conspicuous down to V = 18.0, where they merge
with later spectral type stars (A to F); 
\item $\bullet$ downward V = 15 a group of stars of spectral type F to G starts to become
important. They has the typical bell shape already found in similar diagrams
for field in the 3GQ (Carraro et al. 2008); 
\end{description}

\noindent
We could isolate 80 stars earlier than A0, for which we could assign a spectral type.
Individual estimates of intrinsic colors, reddening, spectral type and distance are listed
in Table~3, together with star IDs, magnitudes and colors.\\
Besides, in Fig.8, top panel, the run of reddening as a function of distance is shown.
Interestingly, the reddening toward VdB-Hagen~4 does not vary much with distance. It gets
to its mean value (0.40$\pm$0.07 mag) very close to the Sun, and at larger distances, up to almost 20 kpc,
keeps basically  constant.\\

\noindent
Apart from the obvious central concentration produced by the star cluster VdB-Hagen~04 (see again Fig~.3)
, these early type stars are evenly distributed across the field implying that the field
itself has a clear young component at these large distances.

\section{The field toward Ruprecht~30}

\noindent
{\bf Field 2}:

The corresponding CMD and the TCD for Field 2 are shown in the right panel
of Fig.~5, and in Fig.~7.\\

\noindent
The {\it blue plume} in the CMD looks quite different from the one in Fig.~3, left panel, for VdB-Hagen~4; 
it shows a larger color spread, it spans a larger range in magnitude and also extends to fainter magnitudes. 

The inspection of Fig.~7 allows us to suggest the following:

\begin{description}
\item $\bullet$ no stars earlier than A0 can be found for V brighter than 12.0 mag;
\item $\bullet$ a first clear group of early spectral type appears at V = 13.0, while a second one
is visible at V = 16.0; 
\item $\bullet$ in-between, and at V fainter than 16.0, there are several late B and early A stars almost
everywhere, which then disappear completely at V larger than 18.0, where they are mixed with late
A and F stars;
\end{description}

We counted 200 stars earlier than A0 in this field, for which we could assign a spectral type.
Individual estimates of intrinsic colors, reddening, spectral type and distance are listed
in Table~4, together with star IDs, magnitudes and colors.\\
The middle panel of Fig.~8 shows the run of the reddening as a function of magnitude. As in
the case of VdB-Hagen~04, almost all the reddening accumulates close to the Sun, and then it
increases very smoothly with distance up to 20 kpc. 
The mean reddening is 0.466$\pm$0.117, and occurs mostly within 5 kpc from the Sun.\\

\noindent
{In the same Fig.~8 we plot in the bottom panel the same run of the reddening as a function of magnitude,
but for the direction centered in the Canis Major overdensity($l=244^{o}$, $b=-8^{o}$), and derived 
by performing exactly the same
analysis as above for the data-set presented in Carraro et al. (2008), properly normalized to take into
account the same area coverage.  
Exactly as in the cases of VdB-Hagen~04 and Ruprecht~30, we find in this direction that the reddening
jumps up immediately in the solar vicinity and keeps almost constant all the way to the limit of our photometric
data-set. No particular stars lumps are evident, on the contrary we find instead stars located almost uniformly 
from the Sun up to at least 15 kpc.}

\section{The spatial distribution of young stars in the outer disk}
The trend of young star density as a function of helio-centric distance is shown in Fig.~9
for VdB-Hagen~04 (lower panel) and Ruprecht~30 (upper panel).
In this figure star counts are expressed in stars per cubic parsec using as distance bin half a kpc.
The logarithmic counts have also shifted by an arbitrary value for the sake of visibility.
In a few distance bin, we did not have any stars and had to interpolate linearly from
neighbor bins.\\

\noindent
The two distributions are similar up to 9 kpc from the Galactic Center, 
then they fall down up to about 13 kpc from Galactic Center.
In this distance range, the star density goes down faster toward  VdB-Hagen~04. This can
be understood since  VdB-Hagen~04 is at higher Galactic latitude than Ruprecht~30.
At about 13 kpc from the Galactic Center the two profiles cross and flatten  up to 16 kpc,
then they keep falling down up to $\sim$22 kpc from the Galactic center, which corresponds to the limit
of our photometry.
The shape of the derived profile can be tentatively interpreted as follows.
Along the two lines of sights and beyond half a kpc from the Sun the density drops
until the Perseus arm is reached. We recall from previous studies of our group (V\'azquez et al. 2008) 
that at the longitudes into consideration the Local arm is not very important and therefore
does not contribute many early type stars beyond 500 pc from the Sun.
The change of slope between 13 and 16 kpc from the Galactic Center probably indicates the presence and size of the
Perseus arm in this region of the third Galactic quadrant, where the Galactic warp
reaches its maximum height below the formal $b=0^{o}$ Galactic plane (Moitinho et al. 2006).
Beyond 16 kpc from the Sun, we enter an almost empty region until the outer arm (Norma-Cygnus) is reached.
This arm is quite extended and sparse, and - due to Galactic rotation- is very far
away in this portion of the disk and does not contribute much in terms of young stars in the area
we are probing.
On the overall, however, the trend of star density in the outer disk looks like more
an exponential trend with some structures, than an abrupt cut-off as predicted by models.\\

\noindent
To provide a more quantitative assessment, we fit the OBA star counts found 
in the direction to the regions of Ruprecht~30 
and VdB-Hagen~04 by adopting  a simple exponential law of the 
form:

\begin{equation}
\rho(R,l,b,age) = \rho_0 \cdot e^{-(R-R_0)/H}
\end{equation}

\noindent
where H (the scale length in kpc) 
is a free parameter. The same equation is valid for ages smaller than 100 Myr, typical
for the OBA spectral type stars we are considering here.\\
The galactocentric distance has been computed as 

\begin{equation}
R = \sqrt(R_0^2+d^2 \times cosb^2-2 \times R_0 \cdot d\cdot cosl \cdot cosb) 
\end{equation}

\noindent
and for $R_0$, the Sun distance to the Galaxy 
center, we adopted 8.5 kpc. \\
The normalization parameter to the local density for 
early type stars, $\rho_0$, was taken from Reed (2001)  where 
it was stated that the local density of OB-type stars is $9.12 \cdot 10^{-7}$ stars per $pc^{-3}$ . This last value 
is somehow uncertain given the scarcity of stars of these spectral types 
in the solar neighborhood. \\

Beside, let us mention that Robin \& Cr\'ez\'e (1986a) report in their Table 2 values 
ranging from $0.6 \cdot 10^{-7}$ to $0.5 \cdot 10^{-4}$ for stars between O7 and A0 spectral types. 

At any rate a rough estimate of the local density of OBA stars according to our own star counts 
yields a local density $7.07 \cdot10^{-7}$ stars per $pc^{-3}$, more in line with  Reed (2001) findings.

Three attempts were then made to fit star counts using scale lengths H=1.0, H=1.3 and 
H= 1.5 kpc. These values are adequately inserted between the range of scale lengths 
from 1.0 to 5.5 kpc, for thin disc, computed -e.g.- by Rong et al. (2001).\\

\noindent
In all the three cases, the excess of OBA stars we found in our directions is 
evident and significant, and demonstrate that the Galactic thin disk extends much further than
14 kpc from the Galactic center. Beside, at large distances the thin disk appears as quite a disperse
structure.

\section{Discussion and Conclusions}
We have provided evidences for the existence of young diffuse groups of B stars in the extreme
periphery of the Galactic disc, at galactocentric distances between 14 and 22 kpc.\\

\noindent
The two fields we have analyzed are centered on cataloged open clusters: VdB-Hagen~04 and
Ruprecht~30. However, the most young stars we found are evenly distributed across the field and
have quite a significant distance spread, both facts being incompatible with the presence of
physical star clusters. We found only a marginal concentration in the center of Field~1,
compatible with the small, distant star cluster VdB-Hagen~04 (Carraro \& Costa 2007). \\

\noindent
The results presented here, together with those from other groups (Snell et al. 2002; Yun et al.
2007, 2009;  Brand \& Wouterloot 2007), demonstrates that the Galactic thin disc does not have
a sharp cut-off at R$\sim$14 kpc, on the contrary to what it has been commonly believed, and
that active star forming regions are present in its outer limits. Our results also show that,
as indicated by the values of $Z_{GC}$ given in Table~1, the thin disc bends considerably in
the 3GQ, emphasizing once more the importance of the Galactic warp (Momany et al. 2004, 2006;
Moitinho et al. 2006).\\

\noindent
We recall that the overdersity of stars we found in the outer disk beyond the 
model cut-off is not limited to OAB stars (thin disk), but extend to M  giant stars -in the thick disk- as well,
as recently shown by Momany et al. (2006).

\noindent
Our findings indicate that a major revision of the Galactic models which aim to predict the
stellar population in the outer Galactic disc is required.

\acknowledgments
The authors would like to express their gratitude to the referee for a number
of valuable suggestions, which helped to improve the quality of the paper.
G. Carraro is grateful to K. Janes, Y. Momany, T. Bania  and D. Russeil for their useful
input. This study made use of Simbad and WEBDA data bases.
E. Costa acknowledges support by the Chilean Centro de Astrof\'{\i}sica (FONDAP No. 15010003)
and the Chilean Centro de Excelencia en Astrof\'{\i}sica y Tecnolog\'{\i}as Afines (PFB 06).

\clearpage

\begin{figure}
\epsscale{1.0}
\plotone{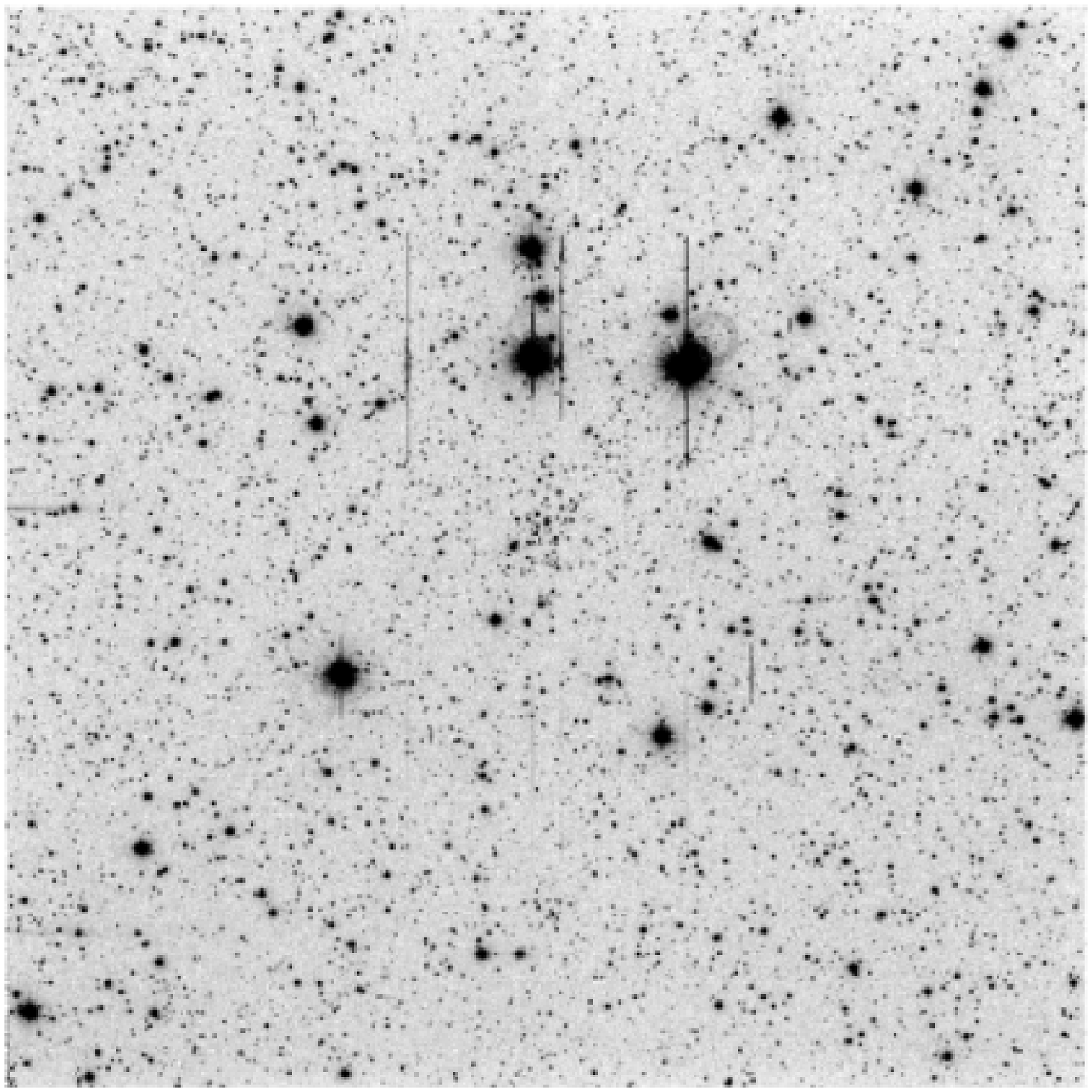}
\plotone{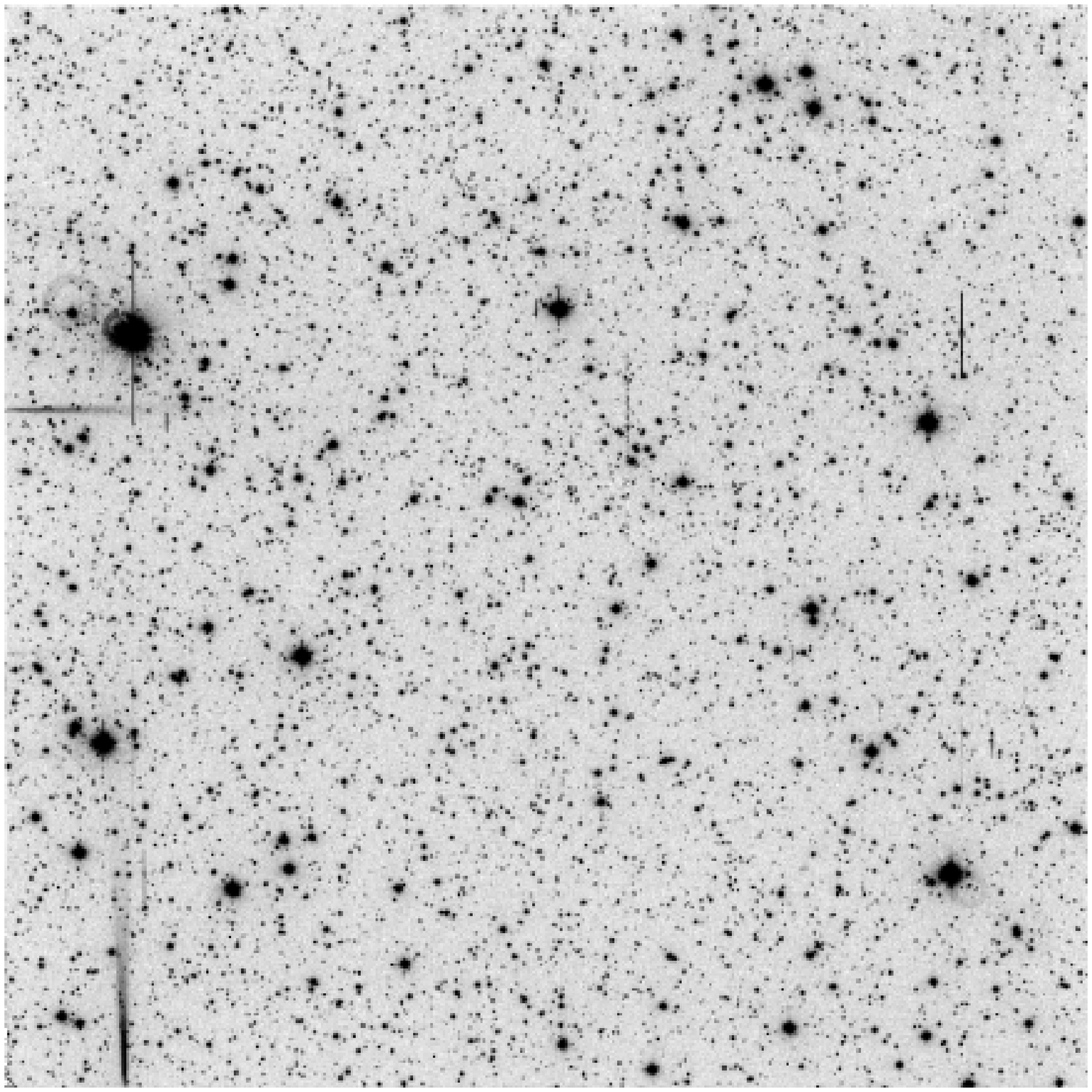}
\caption{A V 600 secs frame in the region of VdB-Hagen~04 (left) and Ruprecht~30
(right). The field of view is 20 arcmin on a side. North is down, East to the right.}
\end{figure}

\clearpage
\begin{figure}
\epsscale{1.0}
\plotone{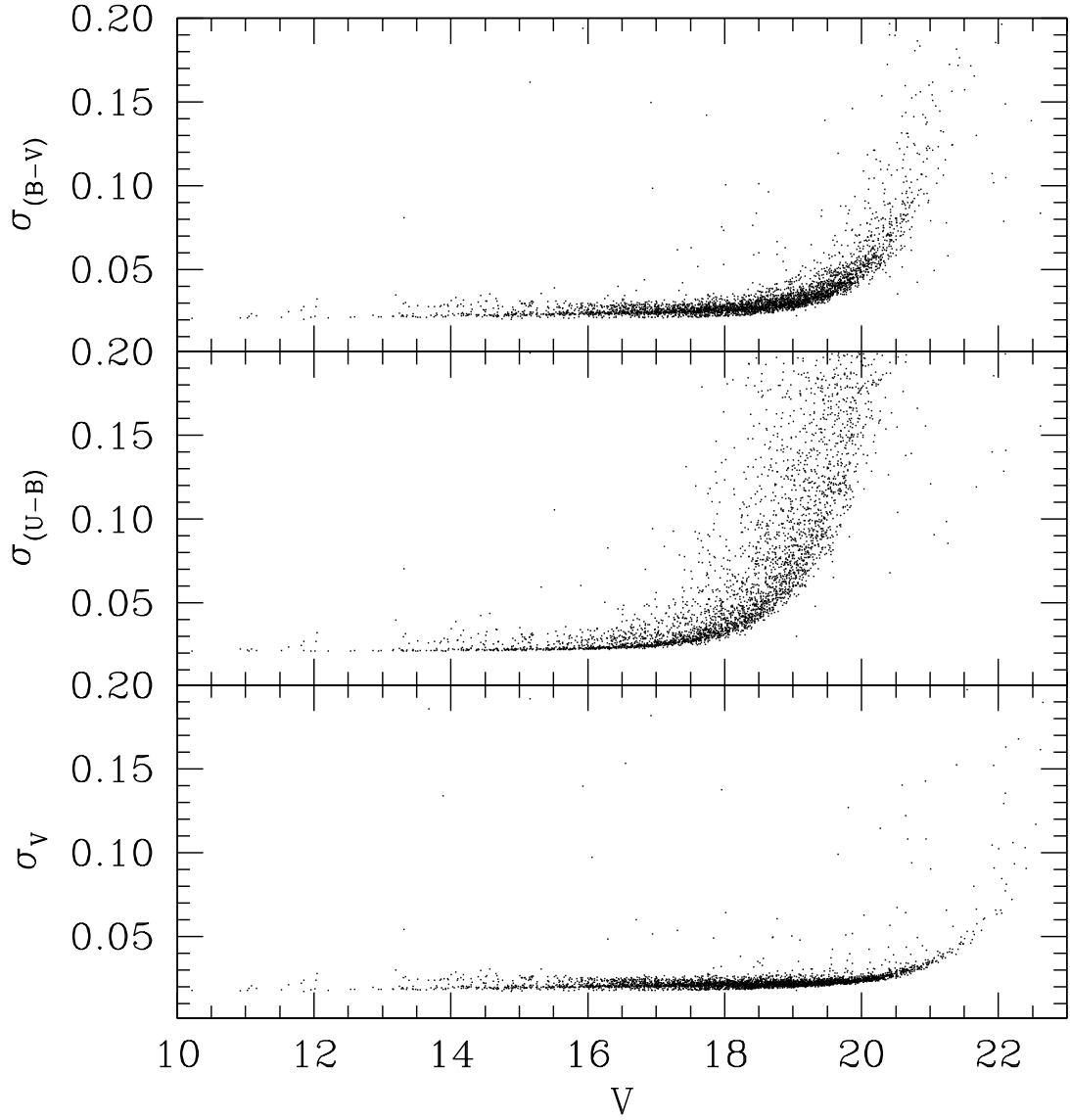}
\caption{Photometric errors in $V$, $(B-V)$, and $(U-B)$ as a function of $V$
magnitude.}
\end{figure}

\clearpage
\begin{figure}
\epsscale{1.0}
\plotone{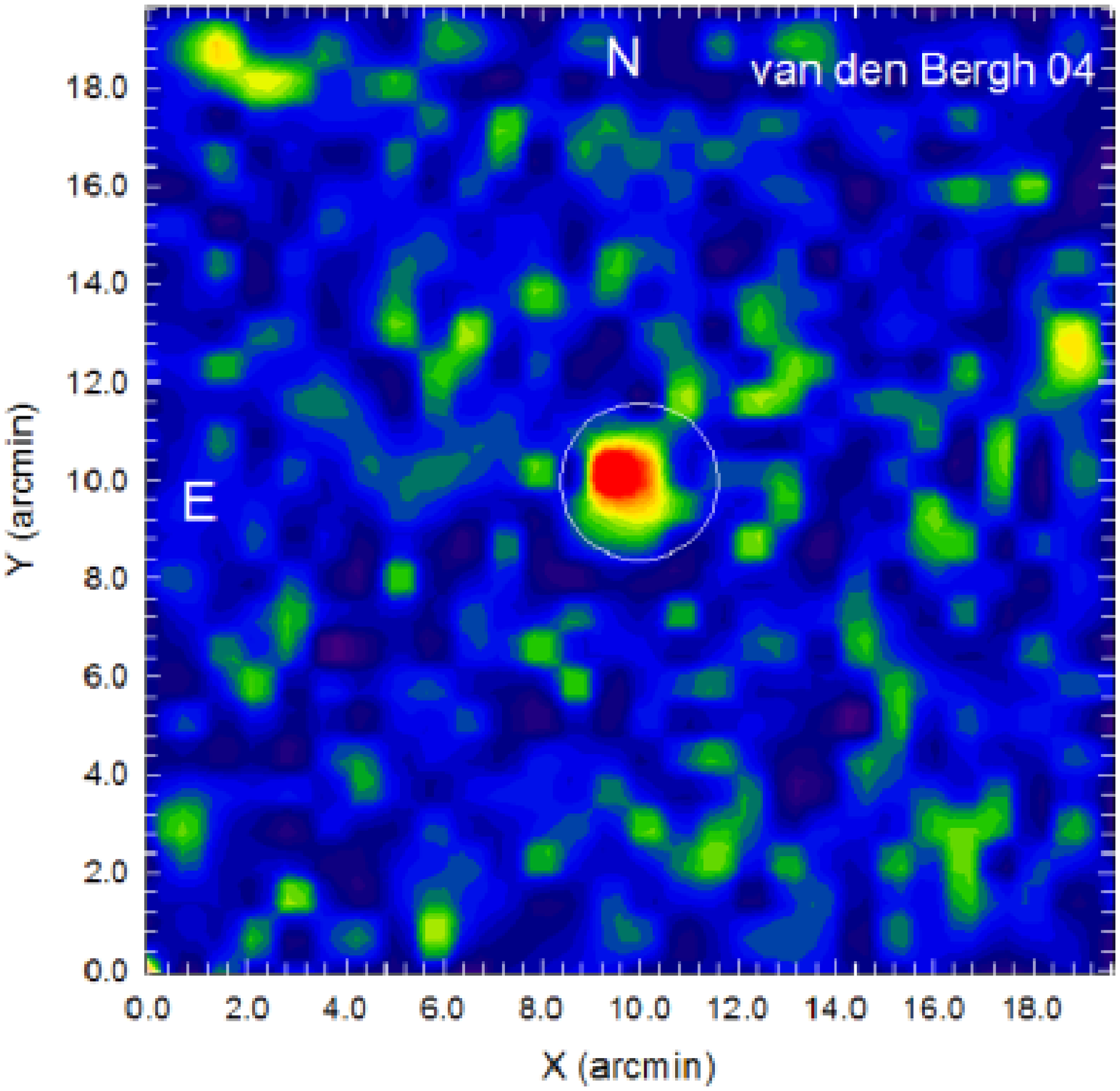}
\plotone{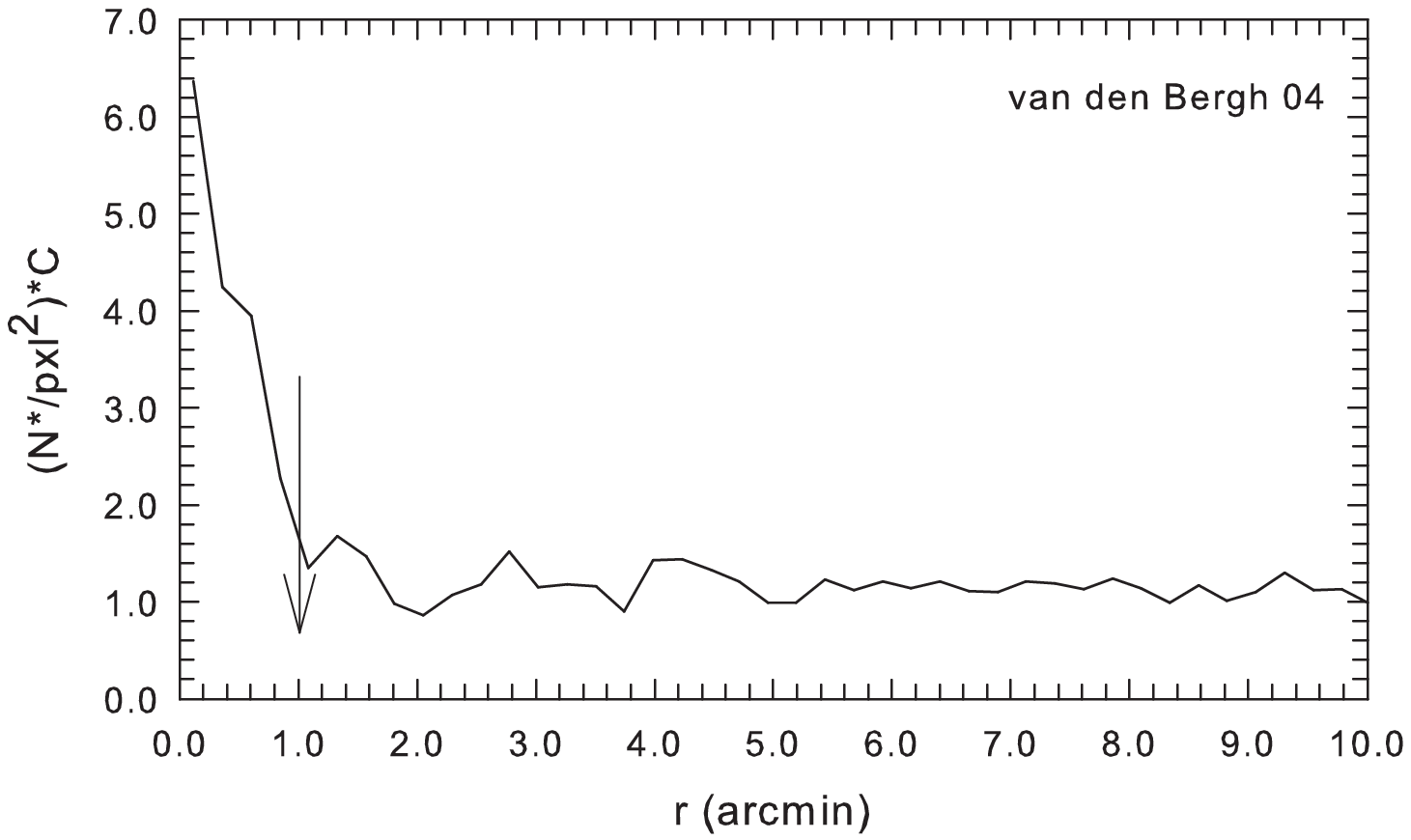}
\caption{Density map (left) and surface density profile (right) in the area
of VdB-Hagen~04}
\end{figure}

\clearpage
\begin{figure}
\epsscale{1.0}
\plotone{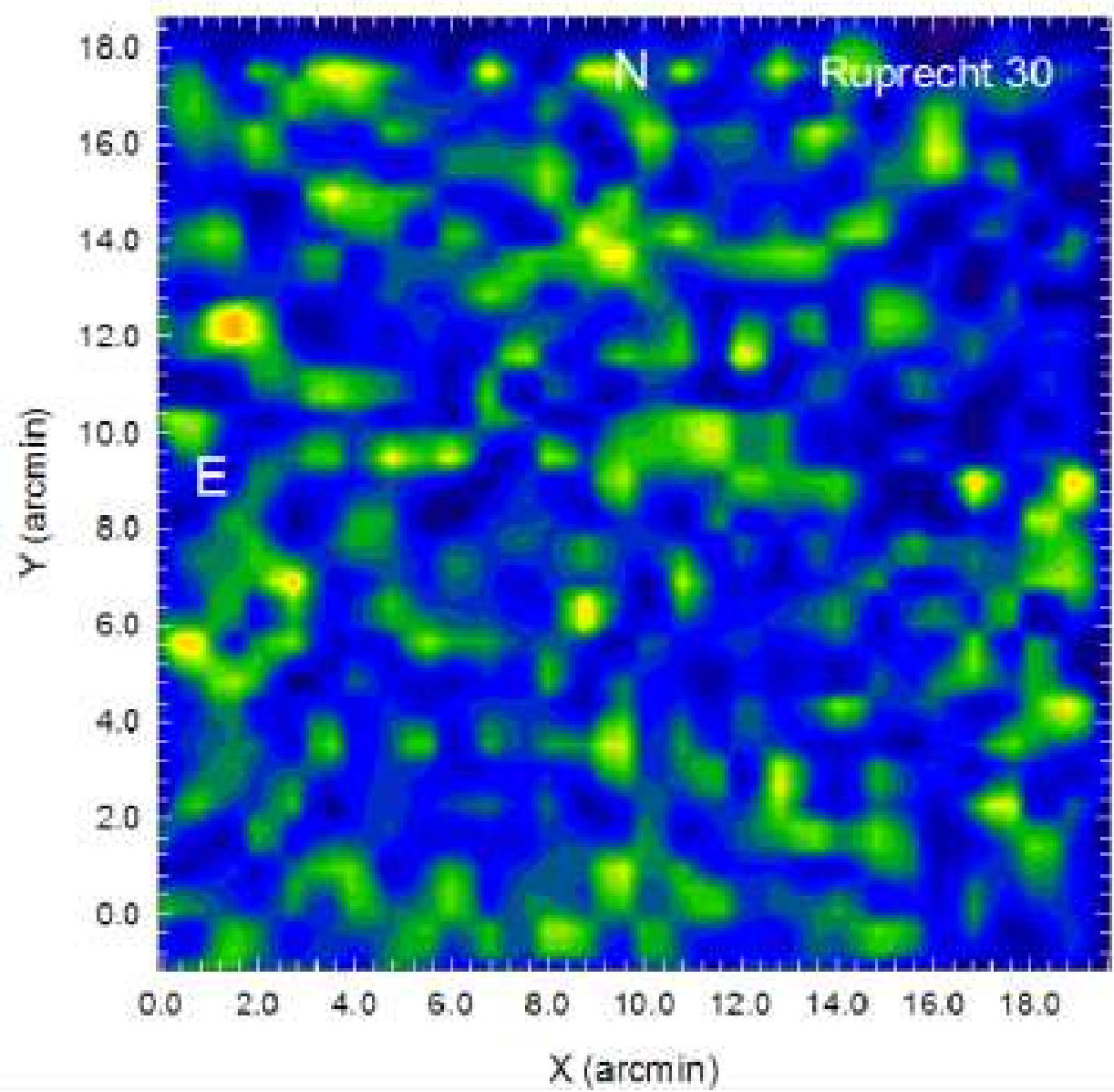}
\caption{Density map in the area
of Ruprecht~30.}
\end{figure}

\clearpage

\begin{figure*}
\epsscale{1.0}
\plotone{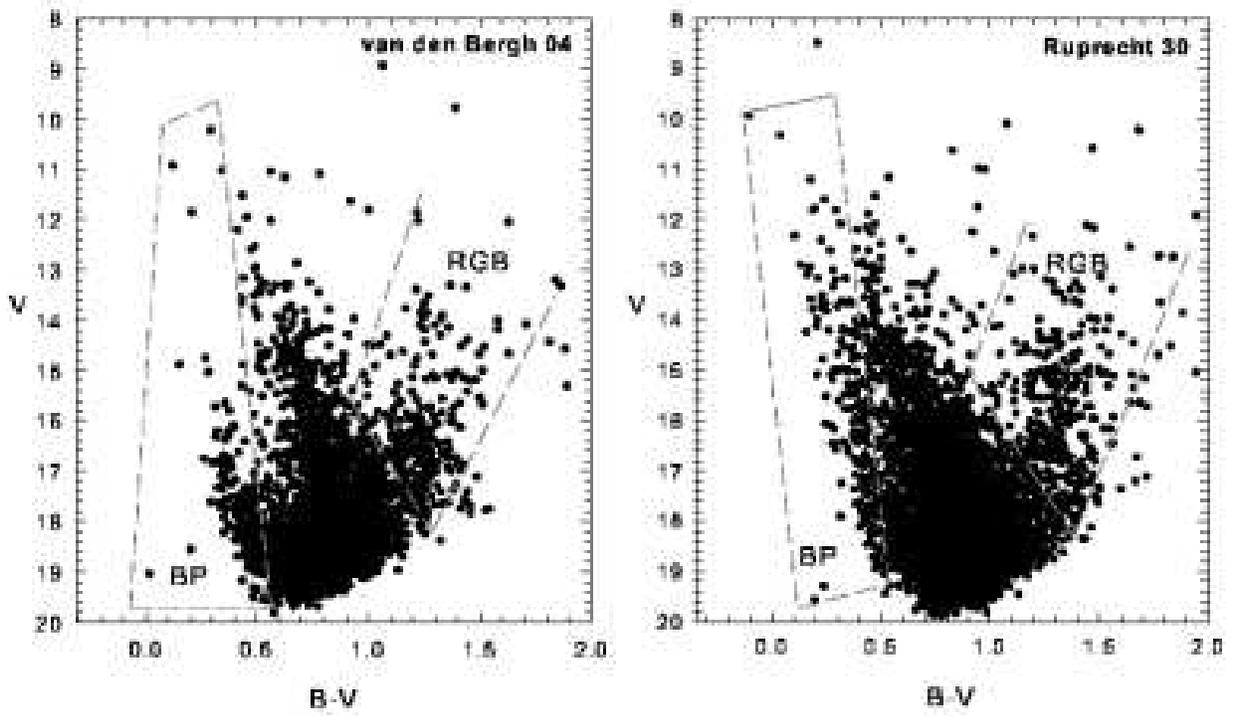}
\caption{Color-Magnitude diagrams for Field~1,
centered on Vdb-Hagen~04 n(left panel), and Field~2, centered on Ruprecht~30 (right panel). 
The regions occupied by Blue Plume and Red Giant stars are indicated.See text for more details}
\end{figure*}

\clearpage

\begin{figure*}
\epsscale{1.0}
\plotone{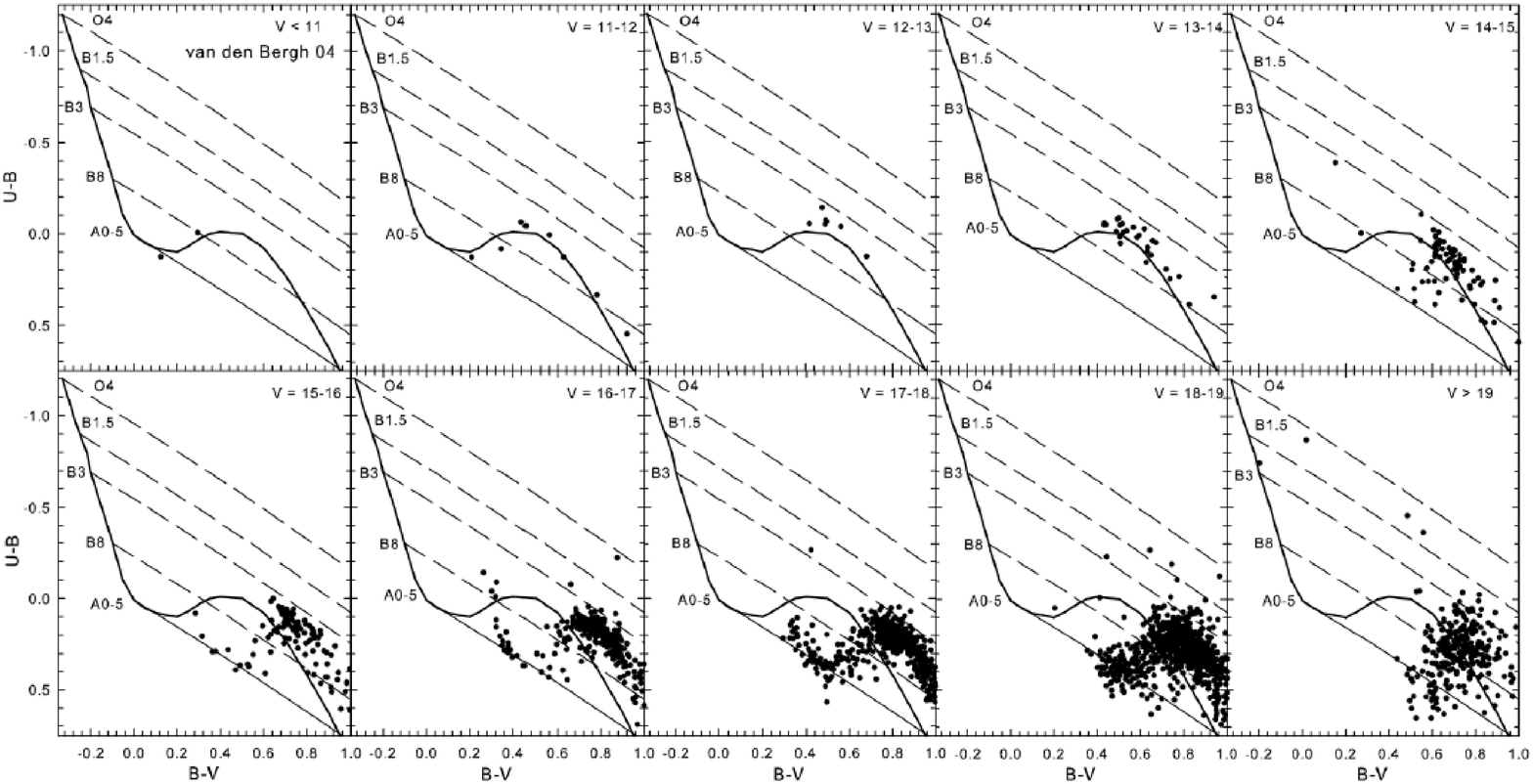}
\caption{TCDs for all the star having UBV photometry in the field of Vdb-Hagen~04, and as function
of the magnitude V. Dashed lines show the run of interstellar reddening for a few typical
spectral types, which are indicated.}
\end{figure*}

\clearpage

\begin{figure*}
\epsscale{1.0}
\plotone{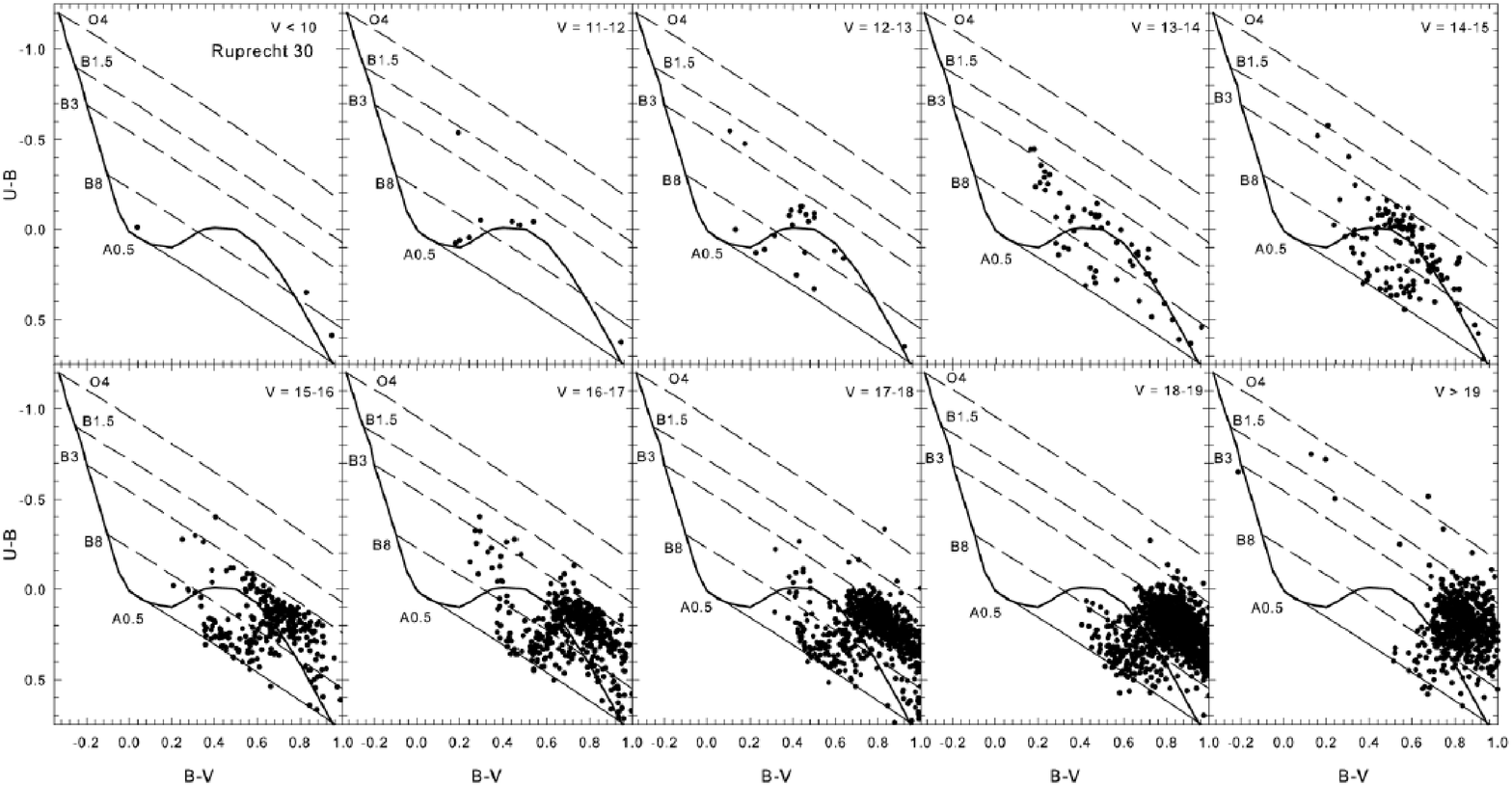}
\caption{TCDs for all the star having UBV photometry in the field of Ruprecht~30, and as function
of the magnitude V. Dashed lines show the run of interstellar reddening for a few typical
spectral types, which are indicated.}
\end{figure*}

\clearpage

\begin{figure}
\epsscale{1.0}
\plotone{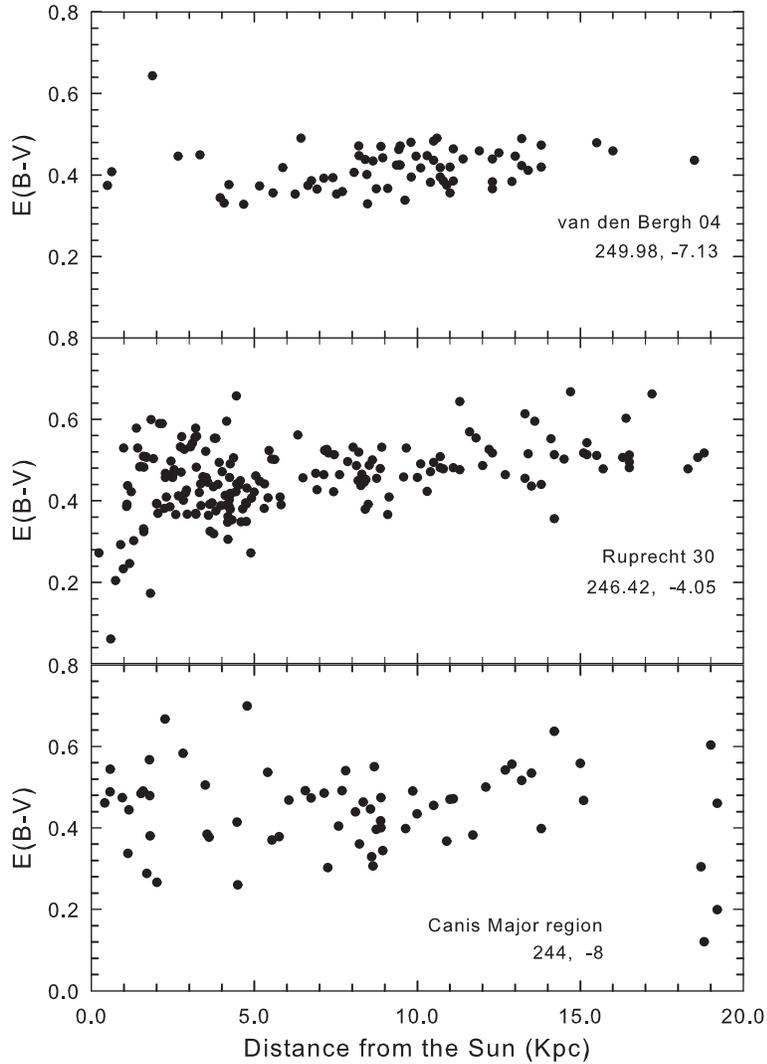}
\caption{Trend of reddening as a function of heliocentric distance along the line of sight to VdB-Hagen~04 (upper panel)
and Ruprecht~30 (middle panel). Lower panel shows the same trend for the Canis Major overdensity direction. Only stars earlier than A0 have been used.
In the upper panel, the compact star cluster  VdB-Hagen~04 is located at a distance of about 12 kpc.}
\end{figure}

\clearpage

\begin{figure}
\epsscale{1.0}
\plotone{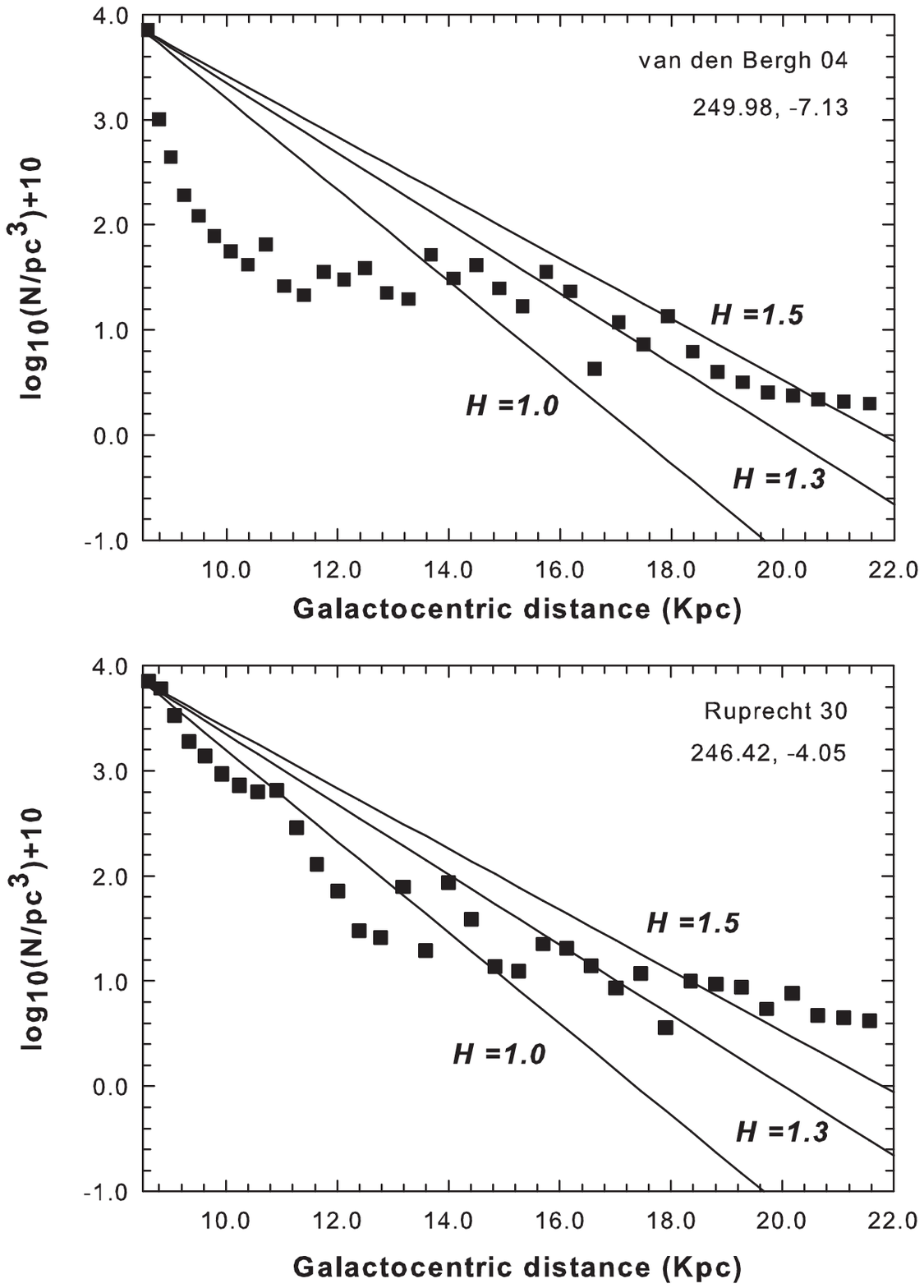}
\caption{Density of stars in the direction of Ruprecht~30 (dashed symbols) and
VdB-Hagen~04 (solid symbols). Density is expressed in star number per cubic parsec.
Only stars earlier than A0 have been considered.}
\end{figure}

\clearpage

 \begin{table*}
 \centering
 \begin{tabular}{ccccccccc}
 \hline\hline
  Field & Designation     & $\alpha (2000.0)$ & $\delta(2000.0)$ & l
  & b &E(B-V)$_{FIRB}$& Constellation\\
     & & & & [deg] & [deg] & [mag] & \\
 \hline
 Field~1& vdB-Hagen~4   & 07:37:44 & -36:04:00  & 249.98 &  -07.13 & 0.49 & Puppis\\
 Field~2& Ruprecht~30   & 07:42:25 & -31:28:00  & 246.42 &  -04.05 & 0.68 & Puppis\\
 \hline\hline
 \end{tabular}
 \caption{Basic properties of the two fields studied in this work.}
 \end{table*}

\clearpage 

\begin{table}
\tabcolsep 0.1truecm
\caption{Log of $BVRI$ photometric observations in the field of VdB-Hagen~04 and Ruprecht~30.}
\begin{tabular}{lcccc}
\hline
\noalign{\smallskip}
Target& Date & Filter & Exposure (sec) & airmass\\
\noalign{\smallskip}
\hline
\noalign{\smallskip}
VdB-Hagen~04 & 2008 January 29 & U & 10, 20, 100, 200, 900, 1800  &1.02$-$1.04\\
             &                 & B & 5, 20, 100, 200, 600, 1800   &1.01$-$1.02\\
             &                 & V & 5, 10, 60, 120, 600, 1200    &1.01$-$1.03\\
SA~98        & 2008 January 29 & U & 2x10, 200, 300, 400          &1.17$-$1.96\\
             &                 & B & 2x10, 100, 2x200             &1.17$-$1.89\\
             &                 & V & 2x10, 100, 2x200             &1.14$-$2.12\\
VdB-Hagen~04 & 2008 February 3 & U & 2x10, 60, 200                 &1.02$-$1.04\\
             &                 & B & 2x10, 30, 60                 &1.01$-$1.02\\
             &                 & V & 2x10, 30, 60                 &1.01$-$1.03\\
SA~98        & 2008 February 3 & U & 2x10, 200, 300, 400          &1.07$-$1.96\\
             &                 & B & 2x10, 100, 2x200             &1.07$-$1.89\\
             &                 & V & 2x10, 100, 2x200             &1.14$-$2.12\\
             &                 & V & 2x10, 100, 2x200             &1.06$-$2.02\\
Ruprecht~30  & 2008 January 31 & U & 5, 20, 100, 200              &1.10$-$1.12\\
             &                 & B & 5, 10, 60, 120               &1.13$-$1.15\\
             &                 & V & 30, 120, 1200                &1.13$-$1.16\\
SA~98        & 2008 January 31 & U & 2x10, 200, 300, 400          &1.15$-$2.20\\
             &                 & B & 2x10, 100, 2x200             &1.05$-$1.79\\
             &                 & V & 2x10, 100, 2x200             &1.05$-$1.79\\
Ruprecht~30  & 2008 February 1 & U & 30, 600, 1500                &1.00$-$1.00\\
             &                 & B & 30, 600, 1500                &1.03$-$1.04\\
             &                 & V & 30, 600, 1200                &1.01$-$1.02\\
SA~98        & 2008 February 1 & U & 2x10, 200, 300, 400          &1.05$-$2.11\\
             &                 & B & 2x10, 100, 2x200             &1.07$-$1.69\\
             &                 & V & 2x10, 100, 2x200             &1.06$-$2.02\\
\noalign{\smallskip}
\hline
\end{tabular}
\end{table}

\clearpage
\begin{deluxetable}{rcrrcccrccc}
\tablecaption{Basic properties of the early spectral type stars encountered in Field~1 (VdB-Hagen~04). The last column indicates the error in distance.}
\tablewidth{0pt}
\tablehead{
\colhead{ID}	& \colhead{V} & \colhead{B-V} & \colhead{U-B} & \colhead{E(B-V)} & \colhead{(B-V)$_0$} & \colhead{(U-B)$_0$} & \colhead{M$_V$} & \colhead{Phot(ST)} & \colhead{Dist(kpc)} & \colhead{$\Delta$ Dist}
} 
\tablecolumns{11} 
\startdata
      1907&       10.22&        0.29&       -0.01&        0.37&       -0.09&       -0.28&        0.62&    B8      &        0.49&        0.02\\
       108&       11.03&        0.35&        0.08&        0.41&       -0.07&       -0.22&        0.80&    B8.5    &        0.62&        0.02\\
      1716&       12.60&        0.48&       -0.14&        0.64&       -0.19&       -0.62&       -0.76&    B3.5    &        1.87&        0.07\\
      2668&       14.89&        0.44&        0.30&        0.45&       -0.01&       -0.02&        1.39&    A0      &        2.66&        0.10\\
      2028&       15.31&        0.43&        0.28&        0.45&       -0.02&       -0.05&        1.31&    A0      &        3.33&        0.13\\
      1377&       14.75&        0.27&        0.00&        0.34&       -0.08&       -0.25&        0.70&    B8.5    &        3.95&        0.15\\
      2116&       15.04&        0.28&        0.08&        0.33&       -0.06&       -0.16&        0.96&    B9      &        4.07&        0.15\\
      2026&       15.80&        0.38&        0.29&        0.38&\phm{-}0.00&\phm{-}0.01&        1.51&    A0.5    &        4.22&        0.16\\
      2434&       15.72&        0.32&        0.21&        0.33&       -0.02&       -0.03&        1.35&    A0      &        4.67&        0.18\\
       118&       16.21&        0.37&        0.28&        0.37&   \phm{-}0.00& \phm{-}0.00&        1.49&    A0.5    &        5.16&        0.20\\
      1582&       16.38&        0.36&        0.28&        0.36&  \phm{-}0.01&  \phm{-}0.02&        1.54&    A0.5    &        5.57&        0.22\\
      1932&       16.63&        0.42&        0.31&        0.42&  \phm{-}0.00& \phm{-}0.00&        1.49&    A0.5    &        5.87&        0.23\\
      1009&       16.21&        0.32&        0.16&        0.35&       -0.04&       -0.10&        1.14&    B9.5    &        6.25&        0.24\\
      3611&       17.02&        0.49&        0.36&        0.49&   \phm{-}0.00&  \phm{-}0.00&        1.46&    A0.5    &        6.43&        0.26\\
      1396&       16.81&        0.38&        0.30&        0.37&   \phm{-}0.01&  \phm{-}0.02&        1.54&    A0.5    &        6.64&        0.26\\
       907&       16.75&        0.38&        0.26&        0.39&       -0.01&       -0.02&        1.41&    A0      &        6.75&        0.27\\
      1975&       16.69&        0.35&        0.23&        0.37&       -0.02&       -0.03&        1.35&    A0      &        6.92&        0.28\\
      1450&       16.93&        0.39&        0.28&        0.39&       -0.01&       -0.01&        1.45&    A0.5    &        7.13&        0.28\\
       418&       16.89&        0.38&        0.24&        0.39&       -0.02&       -0.04&        1.32&    A0      &        7.41&        0.29\\
      2222&       16.94&        0.35&        0.26&        0.35& \phm{-}0.00& \phm{-}0.00&        1.47&    A0.5    &        7.52&        0.29\\
       987&       16.83&        0.34&        0.21&        0.36&       -0.03&       -0.05&        1.28&    A0      &        7.70&        0.30\\
       458&       16.32&        0.32&       -0.01&        0.41&       -0.10&       -0.31&        0.53&    B8      &        8.06&        0.31\\
      1891&       17.44&        0.47&        0.33&        0.47&       -0.01&       -0.02&        1.41&    A0      &        8.20&        0.35\\
      2658&       17.48&        0.45&        0.35&        0.45&  \phm{-}0.01&  \phm{-}0.01&        1.53&    A0.5    &        8.21&        0.33\\
      1043&       17.51&        0.44&        0.34&        0.44&  \phm{-}0.01&   \phm{-}0.01&        1.53&    A0.5    &        8.40&        0.33\\
      1902&       16.98&        0.37&        0.18&        0.40&       -0.04&       -0.11&        1.10&    B9.5    &        8.45&        0.33\\
      1090&       14.88&        0.15&       -0.39&        0.33&       -0.19&       -0.62&       -0.78&    B3.5    &        8.47&        0.31\\
      2866&       17.48&        0.43&        0.31&        0.43&       -0.01&       -0.01&        1.45&    A0.5    &        8.64&        0.34\\
      3431&       16.83&        0.32&        0.12&        0.37&       -0.06&       -0.15&        0.99&    B9      &        8.74&        0.34\\
      1457&       17.71&        0.47&        0.36&        0.47&  \phm{-}0.00& \phm{-}0.01&        1.51&    A0.5    &        8.88&        0.36\\
      3029&       17.17&        0.40&        0.19&        0.44&       -0.05&       -0.13&        1.04&    B9.5    &        8.94&        0.35\\
      1845&       17.10&        0.34&        0.18&        0.37&       -0.04&       -0.09&        1.17&    B9.5    &        9.09&        0.35\\
      2080&       17.52&        0.41&        0.28&        0.42&       -0.02&       -0.04&        1.34&    A0      &        9.36&        0.37\\
      2283&       17.83&        0.47&        0.36&        0.46&  \phm{-}0.01&  \phm{-}0.01&        1.52&    A0.5    &        9.43&        0.38\\
      2038&       17.58&        0.42&        0.29&        0.43&       -0.01&       -0.02&        1.39&    A0      &        9.45&        0.37\\
      2133&       17.66&        0.42&        0.31&        0.42&  \phm{-}0.00& \phm{-}0.00&        1.47&    A0.5    &        9.47&        0.37\\
      1612&       17.75&        0.47&        0.33&        0.47&       -0.01&       -0.02&        1.41&    A0      &        9.47&        0.39\\
      1031&       17.33&        0.33&        0.22&        0.34&       -0.02&       -0.03&        1.36&    A0      &        9.62&        0.37\\
      3885&       17.65&        0.46&        0.28&        0.48&       -0.03&       -0.08&        1.21&    B9.5    &        9.80&        0.40\\
      2616&       17.24&        0.36&        0.16&        0.40&       -0.05&       -0.13&        1.06&    B9.5    &        9.81&        0.38\\
      1970&       17.74&        0.44&        0.30&        0.45&       -0.02&       -0.03&        1.37&    A0      &        9.96&        0.41\\
      3586&       17.16&        0.36&        0.11&        0.42&       -0.07&       -0.20&        0.85&    B9      &       10.10&        0.39\\
      2725&       17.86&        0.44&        0.31&        0.45&       -0.01&       -0.02&        1.40&    A0      &       10.30&        0.42\\
       865&       17.36&        0.35&        0.17&        0.38&       -0.05&       -0.11&        1.09&    B9.5    &       10.40&        0.41\\
      2092&       17.99&        0.44&        0.34&        0.44& \phm{-}0.01& \phm{-}0.01&        1.53&    A0.5    &       10.50&        0.42\\
      1864&       18.02&        0.48&        0.34&        0.48&       -0.01&       -0.01&        1.42&    A0.5    &       10.50&        0.43\\
      3572&       18.01&        0.48&        0.33&        0.49&       -0.02&       -0.03&        1.36&    A0      &       10.60&        0.44\\
      2035&       16.83&        0.30&       -0.04&        0.40&       -0.10&       -0.33&        0.46&    B7.5    &       10.70&        0.80\\
       766&       17.71&        0.40&        0.25&        0.42&       -0.03&       -0.06&        1.26&    A0      &       10.70&        0.42\\
      1346&       17.58&        0.36&        0.21&        0.39&       -0.03&       -0.07&        1.22&    A0      &       10.80&        0.43\\
      1806&       17.76&        0.37&        0.26&        0.38&       -0.01&       -0.02&        1.41&    A0      &       10.90&        0.43\\
      2519&       17.64&        0.34&        0.22&        0.36&       -0.02&       -0.04&        1.33&    A0      &       11.00&        0.45\\
      2139&       18.05&        0.42&        0.33&        0.42& \phm{-}0.01& \phm{-}0.02&        1.54&    A0.5    &       11.00&        0.44\\
      2498&       17.60&        0.36&        0.19&        0.39&       -0.04&       -0.09&        1.17&    B9.5    &       11.10&        0.43\\
      2321&       17.78&        0.43&        0.23&        0.46&       -0.04&       -0.11&        1.11&    B9.5    &       11.10&        0.45\\
      1817&       16.80&        0.32&       -0.09&        0.44&       -0.13&       -0.41&        0.15&    B6.5    &       11.40&        0.44\\
      1162&       18.20&        0.45&        0.32&        0.46&       -0.01&       -0.02&        1.40&    A0      &       11.90&        0.49\\
        29&       16.76&        0.26&       -0.14&        0.38&       -0.13&       -0.42&        0.11&    B6      &       12.30&        0.48\\
      1576&       17.65&        0.33&        0.14&        0.37&       -0.05&       -0.12&        1.07&    B9.5    &       12.30&        0.48\\
      2027&       18.35&        0.44&        0.34&        0.44&  \phm{-}0.01& \phm{-}0.02&        1.54&    A0.5    &       12.30&        0.56\\
       571&       18.33&        0.45&        0.32&        0.45&       -0.01&       -0.01&        1.42&    A0.5    &       12.50&        0.53\\
      1823&       17.72&        0.34&        0.12&        0.38&       -0.06&       -0.16&        0.97&    B9      &       12.90&        0.51\\
      3814&       18.33&        0.44&        0.30&        0.45&       -0.02&       -0.03&        1.38&    A0      &       13.00&        0.55\\
      1446&       17.93&        0.43&        0.15&        0.49&       -0.07&       -0.21&        0.81&    B8.5    &       13.20&        0.53\\
      1565&       18.30&        0.41&        0.29&        0.42&       -0.01&       -0.02&        1.39&    A0      &       13.20&        0.54\\
      1892&       18.39&        0.41&        0.31&        0.41&  \phm{-}0.00& \phm{-}0.00&        1.49&    A0.5    &       13.40&        0.55\\
      2954&       18.14&        0.39&        0.21&        0.42&       -0.04&       -0.10&        1.13&    B9.5    &       13.80&        0.63\\
      2165&       18.39&        0.45&        0.28&        0.47&       -0.03&       -0.07&        1.23&    A0      &       13.80&        0.58\\
       925&       18.68&        0.46&        0.29&        0.48&       -0.03&       -0.07&        1.24&    A0      &       15.50&        0.68\\
      3673&       18.90&        0.46&        0.34&        0.46&       -0.01&   \phm{-}0.00&        1.46&    A0.5    &       16.00&        0.74\\
       937&       19.18&        0.44&        0.33&        0.44&   \phm{-}0.00& \phm{-}0.00&        1.49&    A0.5    &       18.50&        0.85\\
\enddata
\end{deluxetable}

\begin{deluxetable}{rcrrcccrccc}
\tablecaption{Basic properties of the early spectral type stars encountered in Field~2 (Ruprecht~30). The last column indicates the error in distance}
\tablewidth{0pt}
\tablehead{
\colhead{ID} & \colhead{V} & \colhead{B-V} & \colhead{U-B} & \colhead{E(B-V)} & \colhead{(B-V)$_0$} & \colhead{(U-B)$_0$} & \colhead{M$_V$} & \colhead{Phot(ST)} & \colhead{Dist(kpc)} & \colhead{$\Delta$ Dist}
} 
\tablecolumns{11} 
\startdata
       781&        8.49&        0.21&       -0.02&        0.27&       -0.07&       -0.21&        0.81&    B8.5    &        0.23&        0.01\\
      4525&       10.32&        0.04&       -0.01&        0.06&       -0.03&       -0.05&        1.28&    A0      &        0.59&        0.02\\
      5383&        9.93&       -0.10&       -0.19&       -0.05&       -0.06&       -0.15&        0.99&    B9      &        0.66&        0.03\\
      1358&       11.21&        0.18&        0.08&        0.20&       -0.03&       -0.07&        1.23&    A0      &        0.74&        0.03\\
      4683&       11.61&        0.24&        0.05&        0.29&       -0.06&       -0.17&        0.94&    B9      &        0.90&        0.03\\
      4674&       11.79&        0.20&        0.06&        0.23&       -0.04&       -0.11&        1.11&    B9.5    &        0.98&        0.04\\
      1945&       12.28&        0.46&        0.13&        0.53&       -0.08&       -0.26&        0.67&    B8.5    &        0.99&        0.04\\
      1685&       12.08&        0.32&        0.04&        0.39&       -0.08&       -0.25&        0.71&    B8.5    &        1.08&        0.04\\
      4098&       11.82&        0.30&       -0.05&        0.39&       -0.10&       -0.34&        0.43&    B7.5    &        1.09&        0.04\\
      5619&       12.83&        0.42&        0.25&        0.44&       -0.03&       -0.07&        1.24&    A0      &        1.11&        0.04\\
      6027&       12.41&        0.23&        0.13&        0.25&       -0.02&       -0.05&        1.31&    A0      &        1.17&        0.05\\
      1805&       13.22&        0.42&        0.31&        0.42& \phm{-}0.00&  \phm{-}0.00&        1.48&    A0.5    &        1.22&        0.05\\
      1356&       12.62&        0.27&        0.11&        0.30&       -0.04&       -0.11&        1.11&    B9.5    &        1.30&        0.05\\
       431&       12.16&        0.43&       -0.10&        0.58&       -0.16&       -0.53&       -0.34&    B4.5    &        1.38&        0.06\\
      1006&       12.20&        0.39&       -0.11&        0.53&       -0.15&       -0.49&       -0.20&    B5      &        1.42&        0.06\\
      5939&       13.36&        0.44&        0.21&        0.48&       -0.05&       -0.14&        1.01&    B9      &        1.48&        0.06\\
      1950&       13.57&        0.46&        0.27&        0.48&       -0.04&       -0.09&        1.17&    B9.5    &        1.52&        0.06\\
      5258&       12.59&        0.38&       -0.08&        0.51&       -0.14&       -0.45&       -0.01&    B5.5    &        1.60&        0.06\\
       376&       13.00&        0.28&        0.08&        0.33&       -0.06&       -0.16&        0.95&    B9      &        1.60&        0.06\\
      1203&       13.19&        0.29&        0.14&        0.32&       -0.04&       -0.09&        1.16&    B9.5    &        1.61&        0.06\\
      2301&       13.80&        0.47&        0.30&        0.48&       -0.03&       -0.06&        1.27&    A0      &        1.61&        0.07\\
      4915&       12.89&        0.40&       -0.02&        0.51&       -0.12&       -0.40&        0.20&    B6.5    &        1.67&        0.07\\
       657&       13.73&        0.47&        0.23&        0.51&       -0.05&       -0.14&        1.01&    B9      &        1.69&        0.07\\
       480&       12.90&        0.13&        0.00&        0.17&       -0.05&       -0.12&        1.07&    B9.5    &        1.81&        0.07\\
      1216&       12.62&        0.44&       -0.13&        0.60&       -0.17&       -0.57&       -0.54&    B4      &        1.83&        0.07\\
      1814&       13.55&        0.42&        0.09&        0.50&       -0.09&       -0.28&        0.60&    B8      &        1.90&        0.08\\
      4755&       14.17&        0.39&        0.28&        0.39&       -0.01&       -0.01&        1.44&    A0.5    &        2.00&        0.08\\
      4197&       13.65&        0.34&        0.11&        0.39&       -0.06&       -0.18&        0.91&    B9      &        2.01&        0.08\\
      4148&       13.63&        0.32&        0.10&        0.37&       -0.06&       -0.17&        0.94&    B9      &        2.04&        0.08\\
      3135&       12.94&        0.44&       -0.13&        0.59&       -0.17&       -0.56&       -0.50&    B4.5    &        2.09&        0.08\\
      2209&       13.19&        0.45&       -0.09&        0.59&       -0.16&       -0.52&       -0.32&    B4.5    &        2.17&        0.09\\
      3829&       13.33&        0.28&       -0.07&        0.38&       -0.11&       -0.34&        0.40&    B7.5    &        2.24&        0.09\\
      5782&       13.17&        0.34&       -0.12&        0.47&       -0.14&       -0.46&       -0.06&    B5.5    &        2.27&        0.09\\
      1450&       13.66&        0.37&        0.01&        0.46&       -0.10&       -0.33&        0.46&    B7.5    &        2.27&        0.09\\
      3805&       14.35&        0.39&        0.24&        0.41&       -0.03&       -0.06&        1.27&    A0      &        2.30&        0.09\\
      1202&       14.25&        0.36&        0.19&        0.39&       -0.04&       -0.10&        1.15&    B9.5    &        2.41&        0.10\\
      3020&       14.71&        0.48&        0.30&        0.50&       -0.03&       -0.07&        1.24&    A0      &        2.44&        0.10\\
      2185&       13.67&        0.35&       -0.05&        0.46&       -0.12&       -0.38&        0.27&    B7      &        2.49&        0.10\\
      2617&       14.01&        0.38&        0.05&        0.47&       -0.09&       -0.30&        0.56&    B8      &        2.51&        0.10\\
      5869&       13.62&        0.36&       -0.07&        0.48&       -0.13&       -0.42&        0.12&    B6      &        2.54&        0.10\\
       647&       14.23&        0.33&        0.13&        0.37&       -0.05&       -0.13&        1.03&    B9      &        2.59&        0.10\\
       187&       14.73&        0.40&        0.26&        0.41&       -0.02&       -0.04&        1.32&    A0      &        2.67&        0.11\\
      5592&       14.14&        0.43&        0.02&        0.53&       -0.11&       -0.37&        0.31&    B7      &        2.73&        0.11\\
       757&       14.73&        0.44&        0.23&        0.47&       -0.05&       -0.12&        1.08&    B9.5    &        2.75&        0.11\\
       263&       14.80&        0.50&        0.22&        0.56&       -0.07&       -0.19&        0.86&    B9      &        2.77&        0.11\\
      2662&       14.16&        0.33&        0.03&        0.40&       -0.08&       -0.26&        0.67&    B8.5    &        2.82&        0.11\\
      2766&       14.83&        0.48&        0.21&        0.53&       -0.06&       -0.17&        0.92&    B9      &        2.85&        0.12\\
      5340&       14.54&        0.37&        0.14&        0.42&       -0.06&       -0.17&        0.94&    B9      &        2.89&        0.11\\
      4645&       11.81&        0.19&       -0.54&        0.43&       -0.24&       -0.85&       -1.84&    B2      &        2.92&        0.11\\
      3269&       14.43&        0.32&        0.10&        0.37&       -0.06&       -0.16&        0.95&    B9      &        2.94&        0.12\\
      4489&       14.06&        0.41&       -0.06&        0.53&       -0.14&       -0.45&        0.00&    B5.5    &        3.04&        0.12\\
      5642&       14.32&        0.43&        0.00&        0.54&       -0.12&       -0.40&        0.19&    B6.5    &        3.10&        0.12\\
      4495&       14.35&        0.44&       -0.01&        0.56&       -0.13&       -0.42&        0.11&    B6      &        3.18&        0.13\\
      2678&       13.52&        0.23&       -0.22&        0.37&       -0.15&       -0.48&       -0.16&    B5      &        3.20&        0.12\\
      2846&       14.07&        0.44&       -0.08&        0.58&       -0.16&       -0.51&       -0.25&    B5      &        3.20&        0.13\\
      5539&       15.01&        0.35&        0.23&        0.37&       -0.02&       -0.04&        1.34&    A0      &        3.21&        0.13\\
      1969&       15.16&        0.45&        0.25&        0.48&       -0.04&       -0.10&        1.12&    B9.5    &        3.22&        0.13\\
      3561&       13.96&        0.42&       -0.11&        0.56&       -0.16&       -0.52&       -0.31&    B5      &        3.23&        0.13\\
      2035&       15.03&        0.39&        0.21&        0.42&       -0.04&       -0.10&        1.13&    B9.5    &        3.31&        0.13\\
      5306&       14.62&        0.36&        0.05&        0.44&       -0.09&       -0.27&        0.63&    B8      &        3.35&        0.13\\
      3979&       15.28&        0.38&        0.28&        0.39&       -0.01&       -0.01&        1.44&    A0.5    &        3.36&        0.14\\
      4955&       15.17&        0.43&        0.22&        0.46&       -0.05&       -0.12&        1.08&    B9.5    &        3.42&        0.14\\
      2356&       13.73&        0.30&       -0.20&        0.45&       -0.16&       -0.53&       -0.36&    B4.5    &        3.46&        0.15\\
      1654&       15.15&        0.46&        0.17&        0.52&       -0.07&       -0.22&        0.80&    B8.5    &        3.51&        0.14\\
       314&       15.31&        0.43&        0.24&        0.46&       -0.04&       -0.09&        1.16&    B9.5    &        3.52&        0.14\\
      3036&       15.57&        0.44&        0.32&        0.45&       -0.01&       -0.01&        1.43&    A0.5    &        3.57&        0.14\\
      3547&       15.35&        0.36&        0.26&        0.36&       -0.01&       -0.01&        1.44&    A0.5    &        3.59&        0.14\\
      1736&       13.16&        0.21&       -0.36&        0.39&       -0.19&       -0.64&       -0.85&    B3.5    &        3.63&        0.14\\
       405&       12.33&        0.11&       -0.55&        0.33&       -0.22&       -0.78&       -1.49&    B2.5    &        3.64&        0.14\\
      2852&       13.67&        0.24&       -0.25&        0.40&       -0.16&       -0.54&       -0.39&    B4.5    &        3.70&        0.14\\
      4252&       15.67&        0.43&        0.32&        0.43&  \phm{-}0.00& \phm{-}0.00&        1.46&    A0.5    &        3.74&        0.15\\
       786&       13.78&        0.19&       -0.24&        0.32&       -0.15&       -0.47&       -0.08&    B5.5    &        3.75&        0.15\\
      3917&       14.03&        0.39&       -0.17&        0.55&       -0.17&       -0.58&       -0.56&    B4      &        3.77&        0.15\\
      4439&       15.27&        0.35&        0.20&        0.38&       -0.03&       -0.08&        1.20&    B9.5    &        3.82&        0.15\\
      1665&       15.66&        0.52&        0.28&        0.55&       -0.05&       -0.13&        1.03&    B9.5    &        3.82&        0.16\\
      4373&       15.30&        0.40&        0.17&        0.44&       -0.06&       -0.15&        0.99&    B9      &        3.88&        0.16\\
      5985&       15.78&        0.48&        0.31&        0.49&       -0.03&       -0.05&        1.28&    A0      &        3.92&        0.16\\
      4968&       14.21&        0.26&       -0.16&        0.39&       -0.14&       -0.45&        0.01&    B6      &        3.97&        0.15\\
       192&       15.66&        0.45&        0.26&        0.47&       -0.04&       -0.09&        1.18&    B9.5    &        4.01&        0.16\\
      4736&       15.68&        0.40&        0.26&        0.41&       -0.02&       -0.04&        1.34&    A0      &        4.10&        0.16\\
      1513&       13.76&        0.23&       -0.29&        0.39&       -0.17&       -0.57&       -0.53&    B4      &        4.14&        0.16\\
      2045&       14.48&        0.44&       -0.11&        0.60&       -0.17&       -0.55&       -0.46&    B4.5    &        4.15&        0.17\\
      4392&       15.00&        0.29&        0.04&        0.35&       -0.07&       -0.21&        0.81&    B8.5    &        4.18&        0.16\\
      3481&       13.09&        0.16&       -0.44&        0.36&       -0.20&       -0.70&       -1.14&    B3      &        4.19&        0.16\\
       716&       14.78&        0.23&       -0.02&        0.31&       -0.08&       -0.24&        0.72&    B8.5    &        4.19&        0.16\\
      4734&       12.94&        0.18&       -0.47&        0.39&       -0.22&       -0.76&       -1.38&    B2.5    &        4.22&        0.16\\
      2001&       13.65&        0.23&       -0.32&        0.40&       -0.18&       -0.61&       -0.73&    B4      &        4.23&        0.17\\
      5408&       15.66&        0.43&        0.23&        0.46&       -0.04&       -0.11&        1.10&    B9.5    &        4.24&        0.17\\
      3638&       14.98&        0.33&        0.01&        0.42&       -0.09&       -0.30&        0.55&    B8      &        4.25&        0.17\\
      4807&       15.01&        0.31&        0.02&        0.38&       -0.08&       -0.26&        0.68&    B8.5    &        4.26&        0.17\\
      1127&       15.45&        0.38&        0.16&        0.42&       -0.05&       -0.14&        1.01&    B9      &        4.26&        0.19\\
      1355&       15.83&        0.47&        0.27&        0.49&       -0.04&       -0.09&        1.17&    B9.5    &        4.26&        0.17\\
      3938&       15.76&        0.40&        0.26&        0.42&       -0.02&       -0.04&        1.32&    A0      &        4.27&        0.17\\
       798&       13.98&        0.21&       -0.26&        0.35&       -0.16&       -0.51&       -0.29&    B5      &        4.31&        0.17\\
       388&       15.81&        0.47&        0.24&        0.51&       -0.05&       -0.13&        1.05&    B9.5    &        4.36&        0.18\\
      1310&       13.82&        0.25&       -0.30&        0.42&       -0.18&       -0.61&       -0.72&    B4      &        4.43&        0.17\\
      5927&       14.39&        0.48&       -0.16&        0.66&       -0.19&       -0.65&       -0.89&    B3.5    &        4.45&        0.18\\
      1377&       15.64&        0.40&        0.19&        0.44&       -0.05&       -0.14&        1.02&    B9      &        4.46&        0.18\\
      1575&       15.92&        0.42&        0.27&        0.44&       -0.03&       -0.05&        1.29&    A0      &        4.52&        0.18\\
      1952&       15.89&        0.43&        0.25&        0.45&       -0.03&       -0.08&        1.19&    B9.5    &        4.57&        0.19\\
      4478&       15.08&        0.27&        0.00&        0.35&       -0.08&       -0.25&        0.69&    B8.5    &        4.59&        0.18\\
      2685&       15.77&        0.36&        0.22&        0.38&       -0.03&       -0.06&        1.25&    A0      &        4.64&        0.18\\
      2898&       16.01&        0.39&        0.27&        0.40&       -0.01&       -0.02&        1.41&    A0      &        4.73&        0.19\\
       514&       15.18&        0.28&        0.01&        0.35&       -0.08&       -0.25&        0.71&    B8.5    &        4.75&        0.19\\
      2776&       15.47&        0.34&        0.10&        0.39&       -0.07&       -0.19&        0.87&    B9      &        4.75&        0.19\\
      5564&       15.73&        0.39&        0.17&        0.43&       -0.05&       -0.15&        1.00&    B9      &        4.77&        0.19\\
      2010&       15.09&        0.21&       -0.02&        0.27&       -0.07&       -0.22&        0.80&    B8.5    &        4.89&        0.20\\
      1476&       15.94&        0.38&        0.23&        0.41&       -0.03&       -0.07&        1.23&    A0      &        4.91&        0.20\\
      5669&       16.30&        0.42&        0.32&        0.42& \phm{-}0.00& \phm{-}0.01&        1.50&    A0.5    &        4.99&        0.20\\
      1896&       16.10&        0.43&        0.25&        0.46&       -0.04&       -0.09&        1.16&    B9.5    &        5.04&        0.20\\
      4896&       16.21&        0.43&        0.27&        0.45&       -0.03&       -0.06&        1.26&    A0      &        5.16&        0.22\\
      4045&       13.59&        0.18&       -0.44&        0.38&       -0.21&       -0.72&       -1.22&    B3      &        5.30&        0.21\\
      3317&       16.33&        0.43&        0.29&        0.44&       -0.02&       -0.04&        1.34&    A0      &        5.31&        0.22\\
       356&       16.03&        0.37&        0.19&        0.41&       -0.04&       -0.11&        1.10&    B9.5    &        5.42&        0.22\\
       901&       16.36&        0.49&        0.26&        0.52&       -0.05&       -0.13&        1.05&    B9.5    &        5.45&        0.22\\
      4618&       14.53&        0.33&       -0.25&        0.50&       -0.18&       -0.62&       -0.74&    B4      &        5.53&        0.22\\
       217&       15.47&        0.39&       -0.04&        0.50&       -0.13&       -0.41&        0.17&    B6.5    &        5.62&        0.23\\
      1345&       15.77&        0.34&        0.05&        0.41&       -0.08&       -0.25&        0.69&    B8.5    &        5.79&        0.23\\
      2084&       16.34&        0.38&        0.24&        0.39&       -0.02&       -0.05&        1.31&    A0      &        5.82&        0.24\\
      1822&       16.48&        0.50&        0.18&        0.56&       -0.08&       -0.24&        0.73&    B8.5    &        6.33&        0.26\\
       383&       16.78&        0.44&        0.29&        0.46&       -0.02&       -0.05&        1.31&    A0      &        6.49&        0.27\\
      2975&       16.21&        0.39&        0.05&        0.47&       -0.09&       -0.29&        0.58&    B8      &        6.88&        0.31\\
      3592&       16.90&        0.42&        0.29&        0.43&       -0.02&       -0.03&        1.38&    A0      &        6.92&        0.28\\
      4959&       17.09&        0.46&        0.32&        0.46&       -0.01&       -0.02&        1.39&    A0      &        7.13&        0.30\\
        57&       15.71&        0.39&       -0.11&        0.52&       -0.15&       -0.49&       -0.18&    B5      &        7.16&        0.29\\
      5397&       16.48&        0.44&        0.10&        0.52&       -0.09&       -0.29&        0.58&    B8      &        7.16&        0.29\\
      5447&       15.04&        0.35&       -0.26&        0.53&       -0.19&       -0.65&       -0.89&    B3.5    &        7.22&        0.29\\
      2547&       16.02&        0.40&       -0.04&        0.52&       -0.13&       -0.42&        0.11&    B6      &        7.27&        0.29\\
      2899&       16.77&        0.39&        0.20&        0.42&       -0.04&       -0.11&        1.11&    B9.5    &        7.43&        0.30\\
      2491&       16.37&        0.42&        0.04&        0.51&       -0.10&       -0.34&        0.42&    B7.5    &        7.45&        0.32\\
      4165&       17.20&        0.45&        0.31&        0.46&       -0.02&       -0.03&        1.35&    A0      &        7.61&        0.32\\
      3597&       16.76&        0.43&        0.13&        0.50&       -0.08&       -0.24&        0.75&    B8.5    &        7.86&        0.33\\
       920&       17.21&        0.50&        0.26&        0.53&       -0.05&       -0.13&        1.04&    B9.5    &        8.03&        0.34\\
        81&       16.25&        0.38&       -0.04&        0.49&       -0.12&       -0.40&        0.20&    B6.5    &        8.13&        0.33\\
      4503&       17.13&        0.42&        0.24&        0.45&       -0.04&       -0.09&        1.17&    B9.5    &        8.18&        0.34\\
      3025&       14.67&        0.30&       -0.40&        0.52&       -0.22&       -0.79&       -1.51&    B2.5    &        8.20&        0.32\\
      5113&       16.76&        0.38&        0.11&        0.44&       -0.07&       -0.21&        0.82&    B8.5    &        8.27&        0.34\\
      3275&       17.24&        0.44&        0.26&        0.47&       -0.03&       -0.08&        1.21&    B9.5    &        8.30&        0.35\\
       537&       17.34&        0.43&        0.29&        0.45&       -0.02&       -0.04&        1.34&    A0      &        8.37&        0.35\\
       188&       14.24&        0.16&       -0.52&        0.38&       -0.22&       -0.80&       -1.56&    B2.5    &        8.40&        0.32\\
      4263&       16.77&        0.39&        0.10&        0.45&       -0.08&       -0.24&        0.75&    B8.5    &        8.42&        0.34\\
      1209&       16.17&        0.29&       -0.08&        0.39&       -0.11&       -0.37&        0.32&    B7      &        8.49&        0.34\\
      5031&       17.16&        0.45&        0.21&        0.49&       -0.05&       -0.15&        1.00&    B9      &        8.52&        0.35\\
        68&       16.79&        0.42&        0.07&        0.50&       -0.09&       -0.30&        0.56&    B8      &        8.62&        0.36\\
      1205&       17.11&        0.41&        0.19&        0.46&       -0.06&       -0.15&        0.99&    B9      &        8.75&        0.36\\
      2765&       17.19&        0.43&        0.19&        0.48&       -0.06&       -0.16&        0.96&    B9      &        8.86&        0.38\\
      3736&       17.09&        0.46&        0.14&        0.53&       -0.08&       -0.25&        0.70&    B8.5    &        8.91&        0.37\\
      1903&       16.06&        0.25&       -0.15&        0.37&       -0.13&       -0.42&        0.13&    B6      &        9.09&        0.36\\
       411&       15.53&        0.25&       -0.28&        0.41&       -0.17&       -0.57&       -0.54&    B4      &        9.13&        0.36\\
      3750&       14.12&        0.21&       -0.58&        0.46&       -0.25&       -0.91&       -2.21&    B1.5    &        9.58&        0.37\\
      3299&       16.28&        0.39&       -0.13&        0.53&       -0.16&       -0.51&       -0.29&    B5      &        9.66&        0.39\\
      4327&       17.68&        0.44&        0.28&        0.46&       -0.03&       -0.06&        1.26&    A0      &       10.00&        0.42\\
      2866&       15.61&        0.31&       -0.30&        0.49&       -0.19&       -0.66&       -0.93&    B3.5    &       10.10&        0.40\\
      1289&       17.36&        0.38&        0.16&        0.42&       -0.06&       -0.15&        0.99&    B9      &       10.30&        0.42\\
      3958&       17.84&        0.46&        0.30&        0.47&       -0.02&       -0.05&        1.29&    A0      &       10.40&        0.44\\
      3174&       17.67&        0.46&        0.22&        0.49&       -0.05&       -0.14&        1.02&    B9      &       10.50&        0.44\\
      2071&       16.85&        0.39&       -0.04&        0.51&       -0.13&       -0.42&        0.13&    B6      &       10.70&        0.44\\
       744&       17.71&        0.45&        0.23&        0.48&       -0.05&       -0.12&        1.07&    B9.5    &       10.70&        0.46\\
       973&       17.08&        0.39&        0.02&        0.48&       -0.10&       -0.33&        0.44&    B7.5    &       10.80&        0.44\\
      2904&       16.63&        0.35&       -0.12&        0.48&       -0.15&       -0.47&       -0.09&    B5.5    &       11.10&        0.45\\
      3426&       15.25&        0.40&       -0.40&        0.64&       -0.24&       -0.88&       -2.01&    B1.5    &       11.30&        0.46\\
      3489&       17.84&        0.44&        0.24&        0.48&       -0.04&       -0.11&        1.10&    B9.5    &       11.30&        0.48\\
      1847&       17.89&        0.51&        0.21&        0.57&       -0.07&       -0.21&        0.81&    B8.5    &       11.60&        0.49\\
      4227&       16.45&        0.39&       -0.18&        0.55&       -0.18&       -0.59&       -0.62&    B4      &       11.80&        0.48\\
      1472&       16.41&        0.33&       -0.21&        0.49&       -0.17&       -0.56&       -0.50&    B4.5    &       12.00&        0.48\\
      1878&       17.87&        0.47&        0.17&        0.53&       -0.07&       -0.21&        0.81&    B8.5    &       12.20&        0.52\\
      3022&       17.39&        0.42&        0.02&        0.52&       -0.11&       -0.36&        0.33&    B7      &       12.30&        0.51\\
      2800&       17.93&        0.42&        0.19&        0.46&       -0.06&       -0.15&        0.98&    B9      &       12.70&        0.54\\
      3969&       16.10&        0.27&       -0.32&        0.46&       -0.19&       -0.66&       -0.93&    B3.5    &       13.30&        0.52\\
      5064&       16.33&        0.42&       -0.26&        0.61&       -0.21&       -0.72&       -1.19&    B3      &       13.30&        0.54\\
      2637&       16.53&        0.35&       -0.23&        0.52&       -0.18&       -0.61&       -0.70&    B4      &       13.40&        0.54\\
      1601&       16.46&        0.28&       -0.25&        0.44&       -0.17&       -0.57&       -0.54&    B4      &       13.50&        0.53\\
      3600&       17.13&        0.45&       -0.10&        0.60&       -0.16&       -0.53&       -0.38&    B4.5    &       13.60&        0.56\\
       857&       18.12&        0.40&        0.20&        0.44&       -0.05&       -0.13&        1.06&    B9.5    &       13.80&        0.59\\
      5052&       17.24&        0.42&       -0.09&        0.55&       -0.15&       -0.50&       -0.22&    B5      &       14.10&        0.59\\
      1740&       17.91&        0.32&        0.13&        0.36&       -0.05&       -0.13&        1.04&    B9.5    &       14.20&        0.61\\
      2833&       18.02&        0.44&        0.11&        0.51&       -0.08&       -0.26&        0.66&    B8.5    &       14.20&        0.61\\
      2956&       17.82&        0.41&        0.04&        0.50&       -0.10&       -0.33&        0.46&    B7.5    &       14.50&        0.61\\
      2498&       16.83&        0.48&       -0.19&        0.67&       -0.20&       -0.69&       -1.07&    B3      &       14.70&        0.65\\
      3331&       17.63&        0.40&       -0.04&        0.52&       -0.13&       -0.42&        0.13&    B6      &       15.10&        0.63\\
      4181&       18.07&        0.45&        0.08&        0.54&       -0.10&       -0.32&        0.48&    B7.5    &       15.20&        0.65\\
       528&       18.49&        0.47&        0.23&        0.51&       -0.06&       -0.15&        0.99&    B9      &       15.20&        0.68\\
      3059&       18.05&        0.43&        0.07&        0.51&       -0.10&       -0.31&        0.51&    B8      &       15.50&        0.66\\
      2986&       18.46&        0.44&        0.21&        0.48&       -0.05&       -0.14&        1.00&    B9      &       15.70&        0.70\\
      3826&       17.66&        0.38&       -0.07&        0.51&       -0.14&       -0.44&        0.03&    B6      &       16.30&        0.68\\
      6107&       17.45&        0.45&       -0.11&        0.60&       -0.17&       -0.56&       -0.49&    B4.5    &       16.40&        1.46\\
      3720&       16.58&        0.29&       -0.32&        0.48&       -0.20&       -0.67&       -1.00&    B3      &       16.50&        0.66\\
      1106&       18.72&        0.46&        0.25&        0.50&       -0.04&       -0.11&        1.10&    B9.5    &       16.50&        0.76\\
       955&       18.77&        0.48&        0.26&        0.51&       -0.04&       -0.11&        1.10&    B9.5    &       16.50&        0.75\\
      3763&       16.80&        0.45&       -0.28&        0.66&       -0.22&       -0.77&       -1.43&    B2.5    &       17.20&        0.75\\
      3641&       17.25&        0.32&       -0.22&        0.48&       -0.17&       -0.57&       -0.54&    B4      &       18.30&        0.80\\
      3148&       16.45&        0.29&       -0.40&        0.51&       -0.22&       -0.78&       -1.46&    B2.5    &       18.60&        0.74\\
       754&       18.95&        0.48&        0.23&        0.52&       -0.06&       -0.15&        0.97&    B9      &       18.80&        0.91\\
\enddata
\end{deluxetable}

\end{document}